# Optical system for extremely large spectroscopic survey telescope

Ding-qiang Su[1,2*†], Hua Bai[2,3*†], Xiangyan Yuan[2,3], and Xiangqun Cui[2,3,4]

[1] *Key Laboratory of the Ministry of Education, School of Astronomy and Space Science, Nanjing University, Nanjing 210023, China;*
[2] *Nanjing Institute of Astronomical Optics and Technology (NIAOT), Chinese Academy of Sciences, Nanjing 210042, China;*
[3] *University of Chinese Academy of Sciences, Nanjing 211135, China;*
[4] *School of Astronomy and Space Science, University of Chinese Academy of Sciences, Beijing 101408, China*

Received November 10, 2023; accepted December 26, 2023; published online March 7, 2024

This article presents research work on a spectroscopic survey telescope. Our idea is as follows: for such a telescope, a pure reflecting optical system is designed, which should have an aperture and a field of view (FOV) both as large as possible and excellent image quality, and then a strip lensm (lens-prism) atmospheric dispersion corrector (S-ADC) is added, only for correcting the atmospheric dispersion. Given the fund limitation and the simplicity of scaling up, some 12-m telescopes are designed as examples. Su, Korsch, and Meinel put forward the four-mirror Nasmyth systems I and II, which are used in this paper. FOVs of 1.5°, 2°, and 2.5° are selected. For all systems, the image qualities are excellent. Because the S-ADC relaxes the optical glass size restriction, this 12-m telescope with a FOV of 2.5° can be magnified in proportion to a 16-m telescope. Its etendue (from French étendue) and focal surface will now be the largest in the world. In such a telescope, a pure reflecting optical system can also be obtained. A subsequent coudé system is designed with excellent image quality.

## 1 Research and development of a spectroscopic survey telescope and its astronomical achievements in China

In China, Shou-guan Wang proposed the development of an extra-large spectroscopic survey telescope with thousands or more optical fibers in the late 1980s. To initiate this project, in 1994, Shou-guan Wang and Ding-qiang Su developed the basic configuration for the large sky area multi-object fiber spectroscopic astronomical telescope (LAMOST). In particular, Su innovatively proposed an active reflecting Schmidt system, which continuously changes the shape of the first reflecting mirror by active optics during the observation process [1-3]. This telescope has a diameter of 4.3 m, an angular field of view (FOV) of 5°, a linear FOV diameter of 1.75 m, and 4000 optical fibers. At the end of the last century, the diameter of a zone with one optical fiber was 24 mm, and currently, the LAMOST team has achieved a zone diameter of ~8 mm or smaller. From its completion in 2008 to the completion of the DESI (Dark Energy Spectroscopic Instrument) project in 2021, LAMOST was the telescope with the greatest number of optical fibers in the world, equipped with 16 low- and medium-dispersion spectrometers. Comparing with existed telescopes, LAMOST has gotten the largest focal surface area, which can accommodate the greatest number of optical fibers in the world. Cui et al. [4] led the construction of LAMOST (Figure 1). It is difficult and challenging to develop a mirror with a continuously changing shape, especially a segmented, thin, active deformable mirror. The success of LAMOST resulted in the mastery of active optics in China and laid the foundation for developing an extremely large telescope. Xing et al. [5] proposed an optical fiber positioning system in which the focal surface is divided into many small zones (with each zone containing one optical fiber positioner) where thousands of optical fibers can be simultaneously positioned. Currently, the optical fiber positioning system of LAMOST is widely used in spectroscopic survey telescopes worldwide. LAMOST officially began surveying the sky in September 2012, mainly for the study of the structure and evolution of our Galaxy and stellar physics. By June 2023, the spectra of more than 22 million celestial bodies were obtained, and by March 2023, astronomers published 1300 scientific research papers using the LAMOST data, with more than 15000 citations.

*Corresponding authors (Ding-qiang Su, email: dqsu@nju.edu.cn; Hua Bai, email: hbai@niaot.ac.cn)
†These authors contributed equally to this work.

To correct LAMOST's atmospheric dispersion, in 2012, Ding-qiang Su, Peng Jia, and Genrong Liu proposed an innovative strip lens-prism (lensm) atmospheric dispersion corrector (S-ADC) [6], which could not only deliver excellent image quality but also overcome the size limitation of lens materials in the telescope. It has become a key component in extremely large spectral survey telescopes.

In 2015, Hua Bai designed the first four-mirror spectroscopic survey telescope optical system, with an aperture of 10 m, f-ratio of 1 (primary mirror), FOV of 1.4°, and f-ratio of 4, for the MaunaKea Spectroscopic Explorer (MSE; an international collaborative project). The detailed data are contained in an email dated April 11, 2015, from Hua Bai to Rick Murowinski, Project Manager for MSE. From 2019 to January 2021, Bai et al. [7] designed and studied four Cassegrain telescopes with different types of correctors for spectroscopic survey. The parameters of these four telescopes are as follows: aperture of 6.5 m, FOV of 3°, wavelength range of 0.365-0.95 μm, a site altitude of 2500 m, maximum zenith distance of 60°, and maximum corrector lens diameter of ⩽1.66 m. The designs included two correctors with four-piece lenses and two correctors with five-piece lenses, with an f-ratio of ~3.7; all four systems achieved an image quality with a maximum 80% encircled energy EE80 of ⩽0.60 arcsec. For these systems, if the three parameters, namely, the wavelength range of 0.365-1.1 μm, site altitude of 4200 m, and maximum zenith distance of 50°, are changed while all other parameters remain constant, an f-ratio of ~3.7 and an image quality with a maximum EE80 of <0.50 arcsec can be achieved. It should be noted that the maximum diameter of corrector lenses remains ⩽1.66 m.

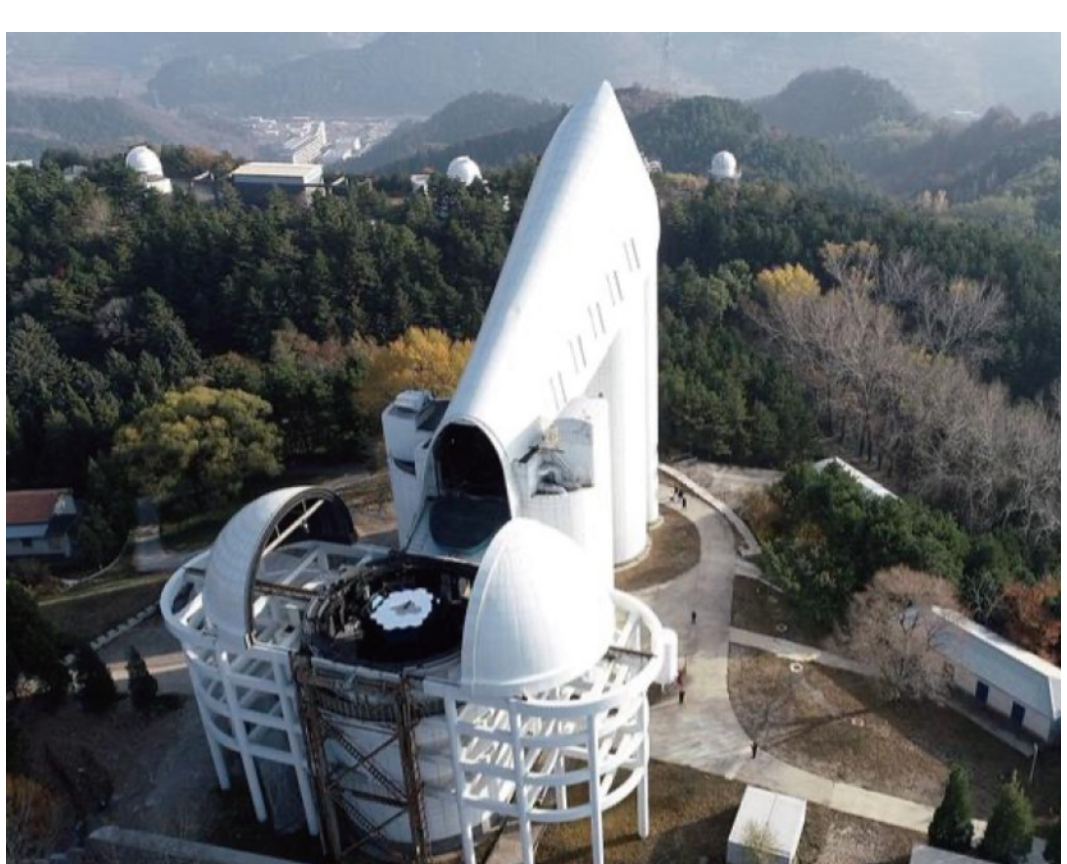

**Figure 1** (Color online) LAMOST at Xinglong Station, National Astronomical Observatory, Chinese Academy of Sciences.

We and our colleagues carried out much other work related to the spectroscopic survey telescope. In 1986, Su [8] first proposed the lensm corrector. Su et al. [9-12] conducted further research. The lensm corrector is one of the basic components of the S-ADC and has been used in spectroscopic survey telescopes such as the 6.5-m telescope [7], WHT [13], and 4MOST. In 2005, Genrong Liu and Xiangyan Yuan proposed and designed several types of small dispersion prisms with different spherical surfaces and glasses, with each prism being used as an optical fiber to correct atmospheric dispersion [14]. Ming Liang proposed a wedged-lens ADC and used it in the DESI optical system [15], and also used in the 6.5-m telescope [7]. Furthermore, the authors and their colleagues carried out significant work on the 12-m general-purpose telescope proposed in China, especially the excellent image quality of the prime focus corrector designed by Su et al. [16-18], which will be used for multi-color photometry and spectroscopic surveys.

In 2015, Su et al. [16-18] proposed the development of a 12-m general-purpose telescope for China. They conducted in-depth research and conducted detailed investigations. The Chinese 12-m general-purpose telescope is also led by Xiangqun Cui.

## 2 From the Chinese 2.16-m telescope's relay mirror to two kinds of Nasmyth systems

### 2.1 Chinese 2.16-m telescope's relay mirror and its significance

The Ritchey-Chrétien (R-C) system is a two nonplane-mirror optical system, which should be free from spherical aberration and should meet the sine condition for all rays at various heights. In the mid-1960s, Ding-qiang Su proposed the idea of using the same secondary mirror for the Cassegrain (R-C) and coudé systems in one telescope. Based on this concept, Ding-qiang Su put forward a series of new coudé systems and several types of relay optical systems. In June 1972, Ding-qiang Su proposed a new coudé system, which only included one relay mirror. This was soon adopted for the Chinese 2.16-m telescope (Figure 2).

The constitutional principle of this system is that from the Cassegrain focus, a concave mirror is used to reimage to the coudé focus. This is also the origin of our three-nonplane-mirror system. As is well-known among experts in the field of optics, third-order spherical aberration, coma, and astigmatism can be eliminated, and when the three-nonplane-mirror system is optimized, excellent image quality can be obtained. When Ding-qiang Su proposed such a coudé system in June 1972, he knew that there were two innovations: (1) the Cassegrain (R-C) and coudé systems shared the same secondary mirror, and (2) a complete excellent coudé system could be obtained when primary, secondary, and relay mirrors were optimized for this coudé system. However, these two innovations could not be simultaneously applied.

Back then in China, all research on the 2.16-m telescope was permitted to be reported and discussed only within the project group due to confidentiality requirements. Thus, our research was not published at that time. A detailed article on the optical system of the 2.16-m telescope was not published until the telescope was successfully assembled in Nanjing and arrived at the Xinglong Station, Beijing Astronomical Observatory (now, National Astronomical Observatories), Chinese Academy of Sciences [19]. Lushun Chen, who was the mechanical structure designer for the 2.16-m telescope coudé system, recorded the following in his work notes in

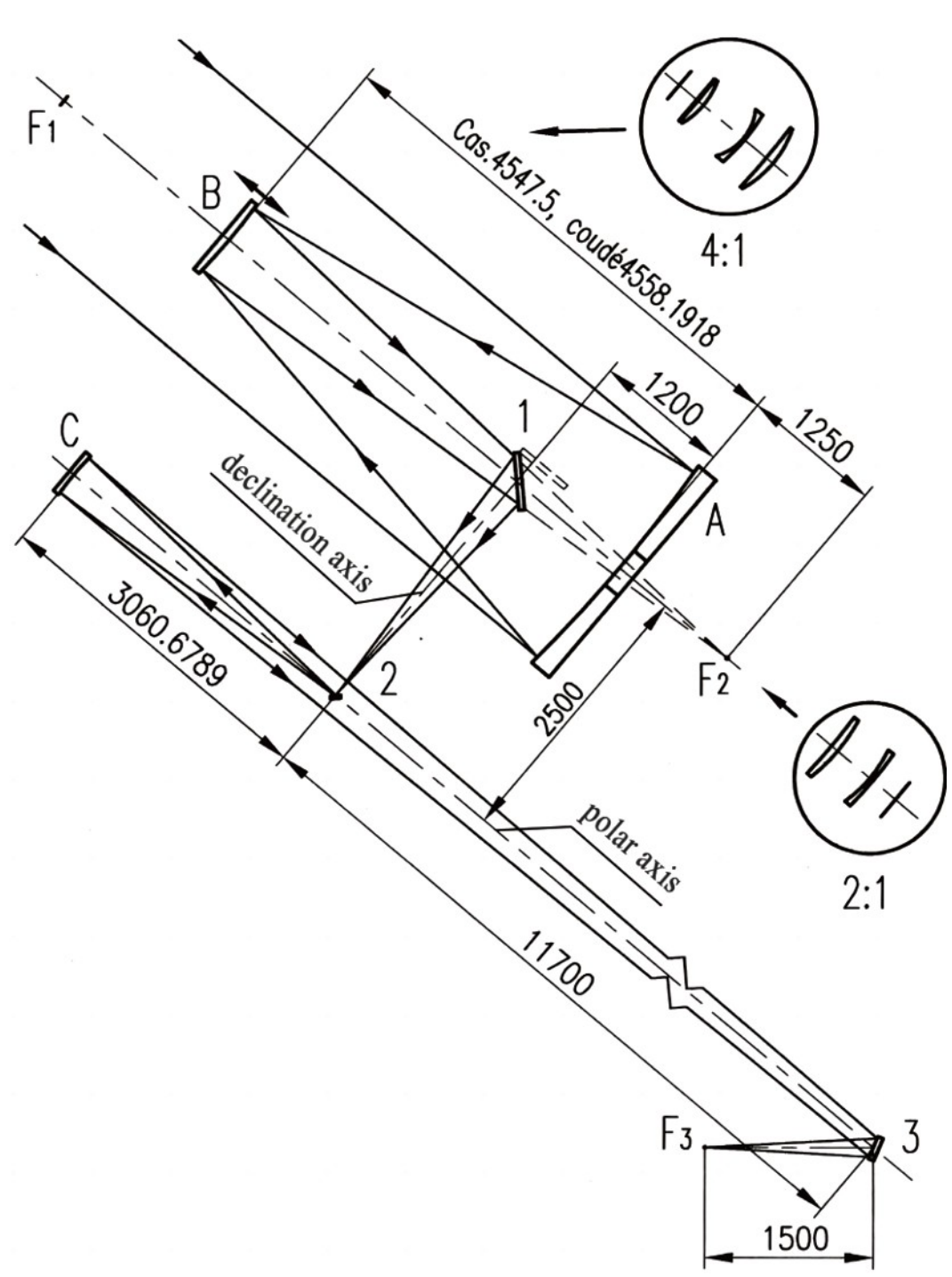

**Figure 2** Optical system of the Chinese 2.16-m telescope. A: primary mirror; B: secondary mirror; C: coudé relay mirror; 1-3: plane mirrors; F1: prime focus; F2: Cassegrain focus; F3: coudé focus. English equatorial mounting. All lengths are in millimeters.

1972: a meeting was held on July 6, 1972, and such a coudé system, which included a relay mirror, was adopted for the 2.16-m telescope. About in 2014, Ding-qiang Su knew the Korsch system [20], which is a three-nonplane-mirror system. Its arrangement is from the Cassegrain focus, and a concave mirror is used to reimage to the last focus. It is similar to our 2.16-m telescope coudé system. The Korsch system was published in *Applied Optics* in December 1972 and was received by this journal on August 10, 1972. Obviously, Ding-qiang Su and Korsch independently proposed the above two similar systems [18]. However, Ding-qiang Su and Korsch are also different: Su's idea came from a desire: the Cassegrain system and the coudé system share the same secondary mirror. Su's ideas can be easily generalized: the Gregorian system and coudé system share the same secondary mirror; the Cassegrain and Nasmyth systems share the same secondary mirror; and the Gregorian and Nasmyth systems share the same secondary mirror. For each of these systems, a relay mirror is added, and then a three-nonplane-mirror system is formed.

In October 1977, a US astronomical delegation, which included 10 distinguished astronomers, visited China. Special permission was granted to us to introduce the above research to them. The idea that both the Cassegrain (R-C) and coudé systems share the same secondary mirror received high praise [21].

As shown in Figures 2 and 3, if only a coudé system is needed, one may raise the following question: why not use a secondary mirror to direct the image to the coudé focus? Our answer is by adding a relay mirror in the coudé system and its image quality can be much better than that of the coudé system with two nonplane mirrors (primary and secondary mirrors). However, in the 2.16-m telescope, the R-C system is predetermined; thus, the shapes of the primary and secondary mirrors are decided. In general, the shape of a relay mirror should be an ellipsoid to eliminate spherical aberration. Even in this situation, we still find the following facts: when exchanging from the R-C system to the coudé system, if the secondary mirror is moved ~11 mm along the chief optical axis and the shape of the relay mirror adopts a suitable oblate, the spherical aberration and coma can be eliminated simultaneously. The image quality of this coudé system will be better than that of the coudé system with two nonplane mirrors [19]. Around 2010, Xiangyan Yuan and Genrong Liu designed the optical system of the Antarctic Kunlun Dark Universe Survey Telescope (KDUST), and Genrong Liu designed one of the optical systems of the Chinese 2-m space telescope. The above two telescope optical systems—Cassegrain and coudé—were proposed by Ding-qiang Su, as shown in Figure 3, in which only the coudé focus was adopted (i.e., the three-nonplane mirrors were optimized only for the coudé focus). Excellent image quality was obtained. In the above two telescopes, the relay mirror is large, and at the telescope tube side, such an arrangement is similar to the four-mirror Nasmyth system II (sect. 2.3).

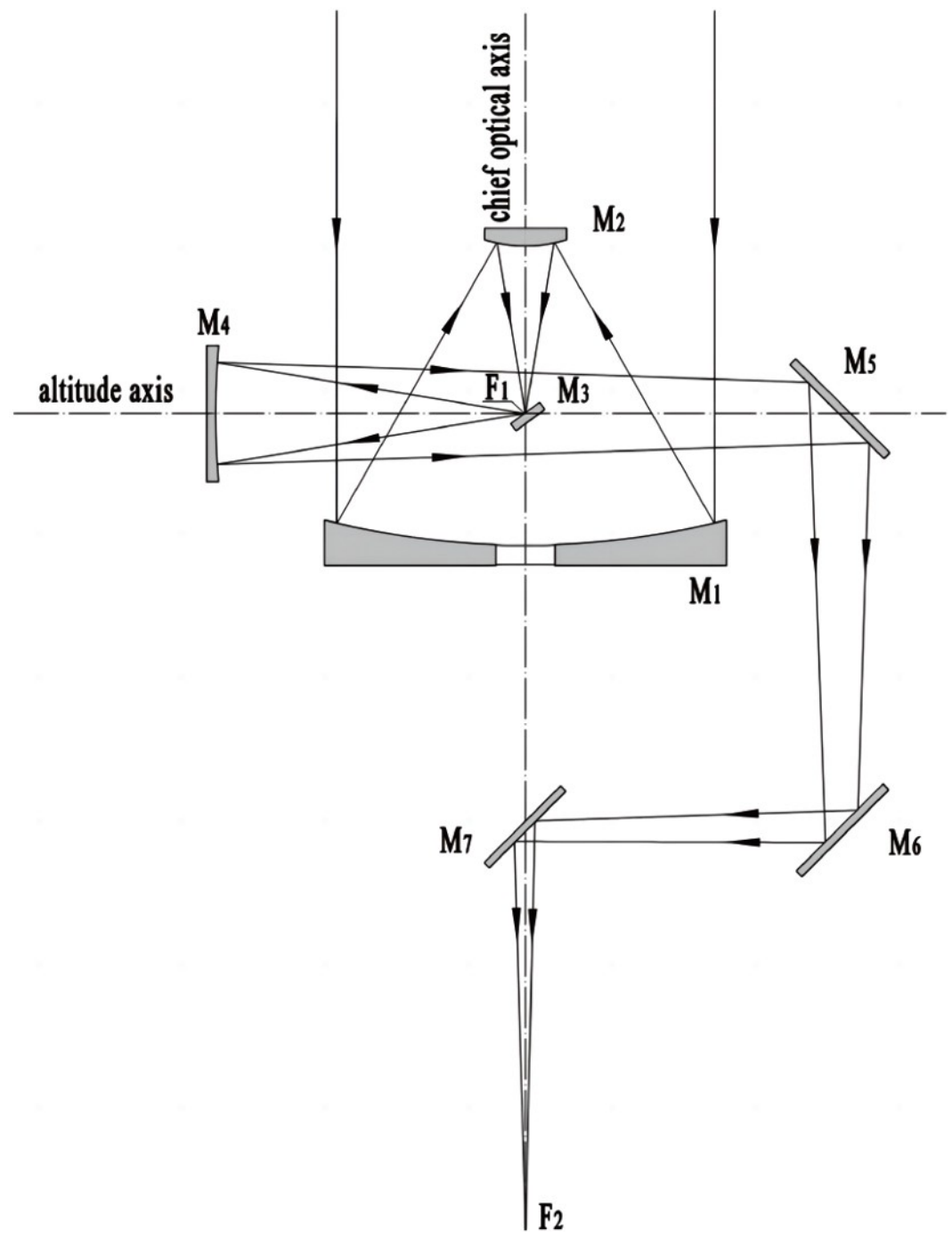

**Figure 3** With a relay mirror, a secondary mirror is shared with a Cassegrain (R-C) system and a coudé system; or we use a complete excellent coudé system with a relay mirror. Alt-azimuth mounting.

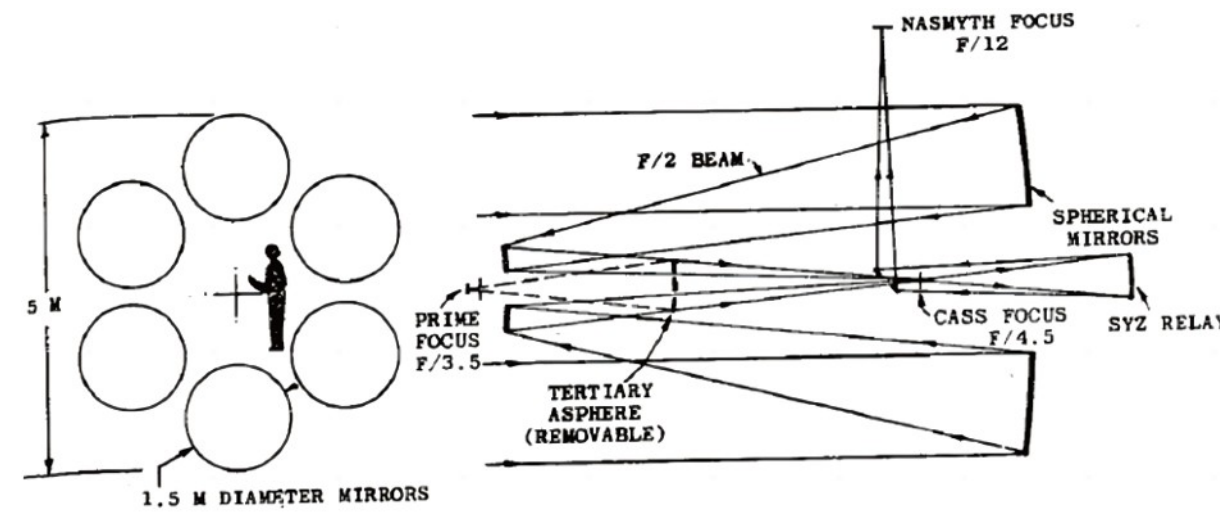

**Figure 4** This is Figure 5 in ref. [24]. The right part of the figure exhibits the extraordinary arrangement of the Nasmyth system proposed by Meinel; sect. 2.2 for details.

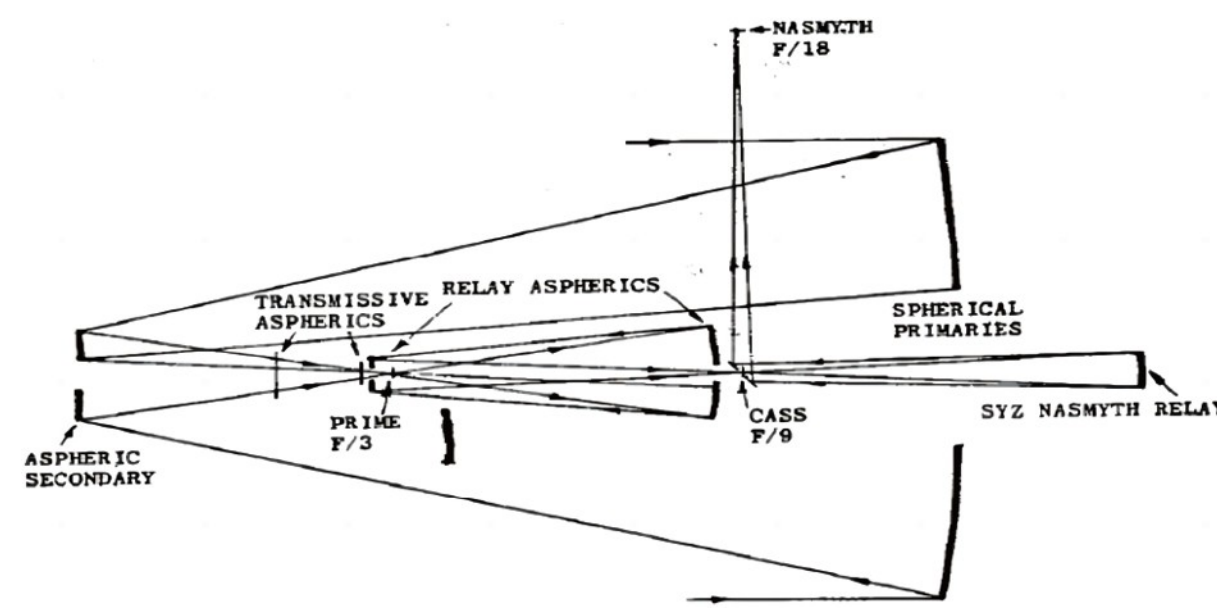

**Figure 5** This is Figure 15 in ref. [24]. The illustration is the same as that in Figure 4.

## 2.2 Four-mirror Nasmyth system I

In 1979, A. B. Meinel and M. P. Meinel, famous astronomers and optical scientists, visited China. They praised the coudé system of the 2.16-m telescope. They named this relay mirror the SYZ relay mirror using the last initials of the designers Ding-qiang Su, Xin-mu Yu, and Bi-fang Zhou. A. B. Meinel and M. P. Meinel adopted the SYZ relay mirror in a number of telescope configurations [22-24]. They not only applied the SYZ relay mirror to the coudé system but were also the first to apply it to the Nasmyth system (Figures 4 and 5). In their arrangement of the Nasmyth system, the SYZ relay mirror is installed on the main optical axis behind the primary mirror, and a 45° plane mirror, which has a hole in its center, is used for the light from the Cassegrain system to pass through the hole to the SYZ relay mirror and then to be reflected by the SYZ relay mirror and the 45° plane mirror to the Nasmyth focus. However, to achieve excellent image quality in Meinel's arrangement, the three nonplane mirrors (primary, secondary, and SYZ relay mirrors) must be optimized. We call such a system the four-mirror Nasmyth system I. The important advantage of the four-mirror Nasmyth system I is that two Nasmyth foci can be used. One can be used for spectroscopic surveys, the other is a pure four-mirror reflecting system, which can be used for refined observation and infrared observation. The shortcoming of the four-mirror Nasmyth system I is that the 45° plane mirror is about 2-3 times larger than the four-mirror Nasmyth system II. Some Chinese astronomical optical experts suggest calling the four-mirror Nasmyth system I the Su-Meinel four-mirror system (or the Su-Meinel optical system).

The 12-m general-purpose telescope being planned in China adopted the four-mirror Nasmyth system I, as shown in Figure 6 and discussed in a few papers [16-18]. In these articles, the structure parameters, image quality, spot dia-grams, and detailed discussions are given.

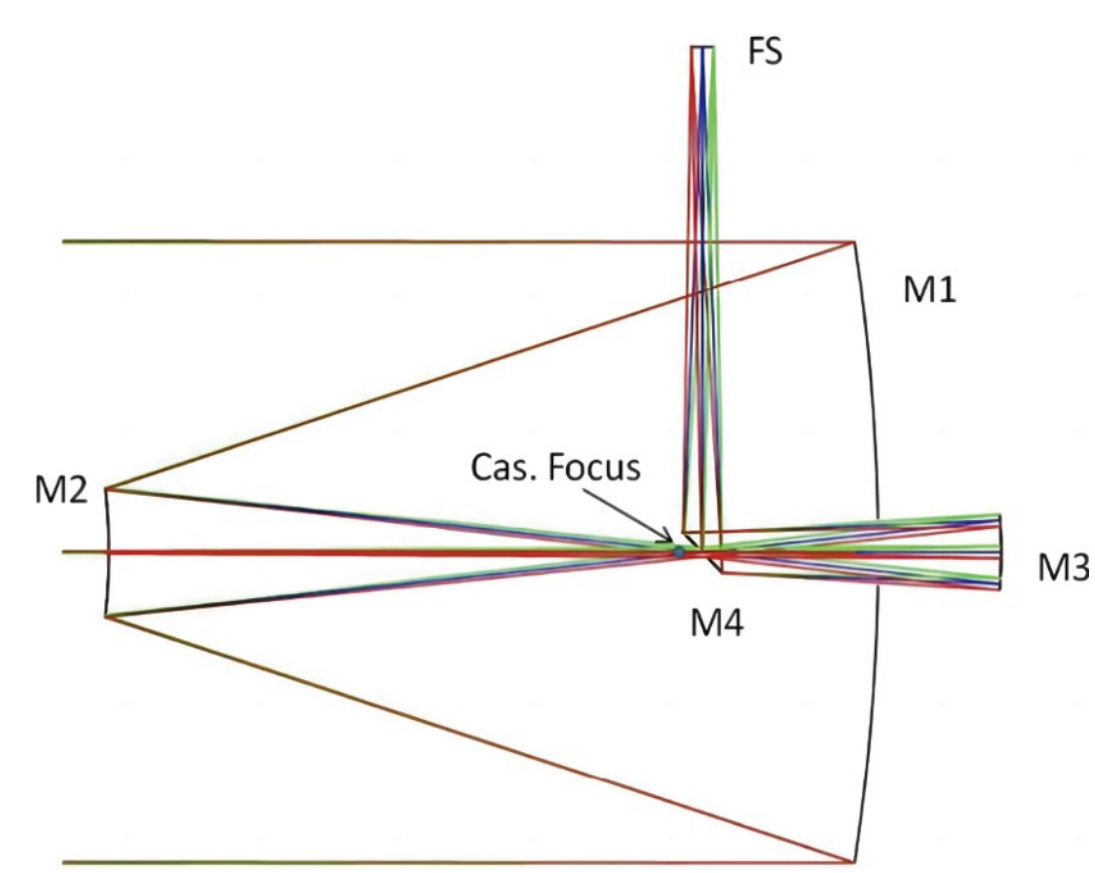

**Figure 6** (Color online) This is Figure 11 from ref. [16]. The Nasmyth system of the 12-m telescope (F/12.8). (M1: primary mirror; M2: secondary mirror; M3: SYZ relay mirror; M4: flat fold mirror).

### 2.3 Four-mirror Nasmyth system II

In Figure 3, if F2 is near M5, the three plane mirrors M5, M6, and M7 can be removed, and Figure 7 can be obtained. Figure 7 shows an optical system that can be treated either as a Nasmyth system and a Cassegrain (R-C) system, which share the same secondary mirror or as a completely excellent Nasmyth system. In 1982, Su et al. [25] proposed an optical system for a 5-m telescope with three systems, namely, "prime", Cassegrain (R-C), and Nasmyth, which share the same secondary mirror. In this telescope, a Nasmyth system, as shown in Figure 7, was adopted. If only the Nasmyth focus is necessary, excellent image quality can be obtained by optimizing the three nonplane mirrors shown in Figure 7. We call such a system the four-mirror Nasmyth system II. The advantage of this system is that it provides a rather small 45° plane mirror that includes only the Cassegrain image. If only one Nasmyth focus is required, such a system is very desirable. The shortcoming of this system is that a second-side Nasmyth focus cannot be used when the relay mirror is large. We found some methods to solve this problem, but they are a little complicated. In April 2015, Hua Bai designed a spectroscopic survey optical system for MSE. In October 2015, we designed such an optical system with an FOV of 2°. Both are four-mirror Nasmyth system II.

The four-mirror Nasmyth systems I and II are very similar: each of them has a relay mirror (SYZ relay mirror), and it is a three-nonplane mirror system. From the arrangement in four-mirror Nasmyth system II, the relay mirror is on one side of the Nasmyth platform, while in four-mirror Nasmyth system I, the relay mirror is on the main optical axis behind the primary mirror. From a four-mirror Nasmyth system I, if we turn a part of the chief optical axis, which includes the relay mirror and the intersection point of the chief optical axis and the altitude axis, around the intersection point onto the altitude axis, and then let the size of 45° plane mirror be changed to the size of intermediate image, the four-mirror Nasmyth system II can be obtained, and *vice versa*. In these two four-mirror Nasmyth systems, except 45° plane mirrors, all mirrors have an unchanged surface shape and spacing distance and provide the same image quality.

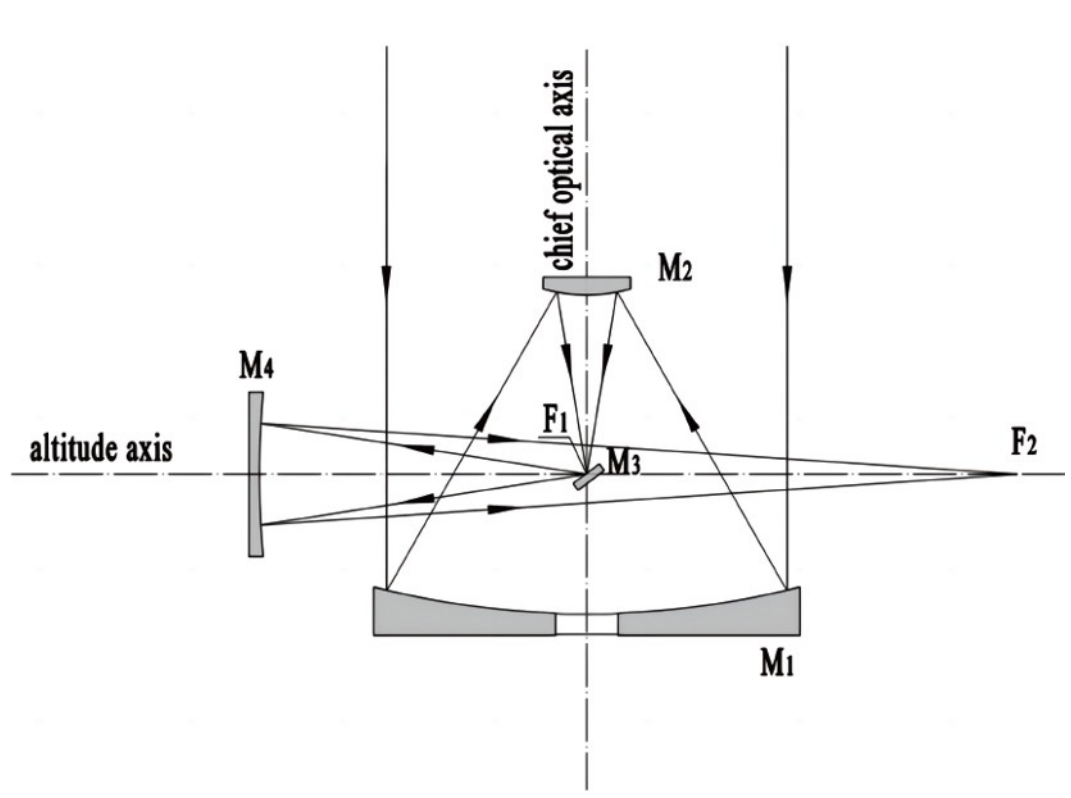

**Figure 7** With a relay mirror, a secondary mirror is shared with a Cassegrain (R-C) system and a Nasmyth system; or we use a complete excellent Nasmyth system with a relay mirror. Alt-azimuth mounting.

The above work on four-mirror Nasmyth systems I and II is based on the original concept of Su et al. [19] (the main content of ref. [19] is introduced by Su's work in the early 1970s) and Meinel and Meinel (1981) [24].

## 3 Discussions of the S-ADC

For visible, near-infrared, and near-ultraviolet light, atmospheric dispersion needs to be corrected. In 2012, Su et al. [6] proposed an innovative S-ADC. It consists of many lensm strips, each with two long wedge lenses, made of two types of glass with a close refractive index but different dispersion [8]. Although its total area is a little larger than the FOV area, this S-ADC consists of many lensm strips; each strip is thin and not very wide in the dispersion direction. Thanks to this new type of ADC, not only the difficulty of obtaining large transparent optical glass can be avoided, but also the aberration caused by this S-ADC is small. By moving such a corrector along the optical axis, different dispersions can be produced, and the atmospheric dispersion at different zenith distances can be compensated.

In this paper, the S-ADC is designed under the condition of no change in the parameters of the original four-mirror reflecting system.

For a point light source (to simplify, "a star" represents "a point light source" below), light arriving at the surface of the S-ADC may have a seam (i.e., a star-illuminated area falls partly on one lensm strip and partly on an adjacent lensm strip). A star-illuminated area has a seam that only corresponds to a straight line on the segmented primary mirror, and these submirrors, which include this straight line, only occupied ~1/10 of the total number of submirrors.

We take the width of the lensm strip slightly more than the diameter of a star-illuminated area on the S-ADC surface when the star is at the required maximum zenith distance of 60°. Thus, for all star-illuminated areas, each does not have or, at most, has only one seam. The atmospheric dispersion is approximately proportional to tan($z$) ($z$ is the zenith distance). The dispersion produced by the S-ADC is directly proportional to the distance from the S-ADC surface to the focal surface. When the zenith distance is from 60° to 30° on the S-ADC surface, the diameter of a star-illuminated area will be reduced to tan30°/tan60° = 1/3, and the probability for a star-illuminated area that has a seam will also be reduced to 1/3.

The S-ADC only is required any star-illuminated area to provide an image that satisfies a standard quality: it enlarges the final image spot by less than 0.10-0.15 arcsec. The diameter of a star-illuminated area on the S-ADC ranges from 184 mm (at $z$ = 60°, sect. 4) to several millimeters (when the star is near the zenith).

Because the dispersion properties between the atmosphere and glass are different, ~1/12 of the compensation residual error remains [6,8]. For example, at a zenith distance of 60°, a wavelength range of 0.36-1.8 μm, and a telescope site altitude of 4200 m, the atmospheric dispersion is 2.8 arcsec, and the compensation residual error is ~0.23 arcsec.

As shown in Figure 8, in this telescope optical system, the S-ADC surface and focal surface are different, and these convex surfaces are placed against each other. Therefore, the distance from the S-ADC surface to the center image on the focal surface and the distance from the S-ADC surface to the image at the edge of the focal surface are different, with the latter exceeding the former by 75 mm (from sect. 4). This value of 75 mm corresponds to a dispersion of 0.31 arcsec; this is the maximum error due to the difference between the S-ADC surface and the focal surface. Obviously, if the dispersion produced by the S-ADC corresponds to the mean surface of the S-ADC surface and focal surface, such an error will be reduced to half, and its maximum is 0.31/2 = 0.155 arcsec.

In the 12-m telescope, the number of S-ADC strips is 7, which can be increased to 9 or 11 (singular numbers are adopted so that the center strip has no seam). Thus, the farthest distance from the S-ADC to the focal surface and the width of the S-ADC strip will be reduced. However, in this case, the dispersion corresponding to 75 mm will increase, as well as the compensation error, but only slightly.

To meet the optical grinding requirements, the edges of the S-ADC strips should be chamfered. All chamfers and sides of the S-ADC strips should be covered with an absorbing black. Hence, the seams will become black belts ~1 mm wide. The thickness of the S-ADC is ~20 mm. The telescope f-ratio is 4. Some oblique rays, which are outside the 1-mm black belt, will also be obstructed. By simple calculation, it is found that the width of an obstructed belt will be ~2.6 mm. For $z$ = 60° and a star at the center of the FOV, the diameter of a star-illuminated area on the S-ADC is 165 mm (from sect. 4). When the obstructed belt is at the center of a 165-mm diameter circle, the obstruction ratio is maximum and is equal to 2.0%. The average obstruction ratio is less than 2.0%. The obstruction ratio is inversely proportional to the diameter of the star-illuminated area, but the probability for the star-illuminated area to meet a seam is directly proportional to the diameter of the star-illuminated area; thus, the average obstruction ratio in all sky areas is unchanged and remains less than 2.0%, which is an acceptable value. Let us discuss the average obstruction ratio in another way: the average illuminance of a large number of different sky areas on the focal surface is a uniform value, on which there are six 2.6-mm-wide blocking strips. Dividing the total area of the six blocking belts by the focal surface area, we obtain an average obstruction ratio of 1.6%. Although the average obstruction ratio is perfectly acceptable, the obstruction ratio is not uniform. For stars near the zenith, the star-illuminated area for each star is very small; if a star meets a seam, the obstruction ratio will be very large, but such stars are only very small in percentage.

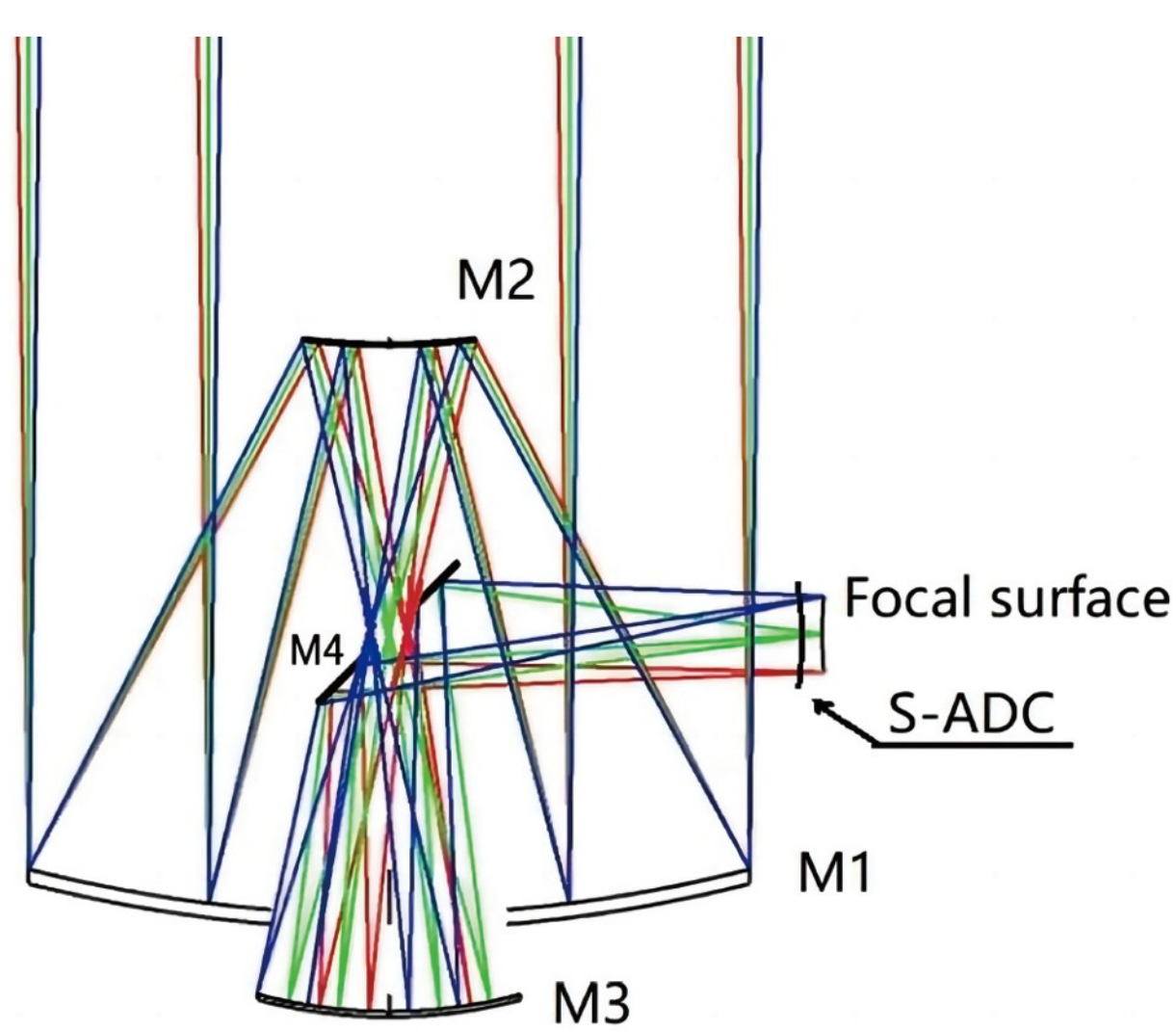

**Figure 8** (Color online) Four-mirror Nasmyth system I for the 12-m spectroscopic survey telescope.

The S-ADC has many advantages: excellent image quality when combined with the four-mirror system, simple construction (only a lensm in each strip), all spherical surfaces, small amount of glass required, and low cost. If the lensms are cemented, only the two outside surfaces produce ghost images. Compared with the corrector (including ADC), the telescope using S-ADC has an outstanding advantage: after moving the S-ADC away from the telescope, it is a pure reflecting telescope optical system with excellent image quality. The use of S-ADC relaxes the limitation of optical glass to the aperture of the telescope so that the aperture of the spectroscopic survey telescope can be 16 m (see below).

The European Southern Observatory (ESO) and others have done a lot of research and planning in the spectroscopic survey and its telescope. Although our idea and design are different from theirs, we also read their articles: "SpecTel: A 10-12 meter class spectroscopic survey telescope" (Astro2020 Decadal Survey facilities white paper, wrote by R. Ellis, K. Dawson, et al.), and refs. [26,27]. Will Saunders and Peter Gillingham designed the optical systems for ESO 11.4-m and MSE 11.25-m spectroscopic survey telescopes that the results are excellent, especially they put forward an innovative idea on ADC, which is a lossless atmospheric dispersion correction [28-30].

## 4 Discussion of some important issues and an example design for a 12-m aperture telescope with an FOV of 1.5°

### 4.1 Discussion of several important issues

(1) Limitations of telescope aperture are as follows.

In this paper, we consider three 12-m telescopes with FOVs of 1.5°, 2°, and 2.5°, respectively, and a width of the strip lensm of the S-ADC ⩽ 200 mm. If these telescopes are scaled up to 20 m, the width of the strip lensm becomes ⩽ 333 mm. There is no problem with the optical glass, so the aperture of these telescopes can be 20 m or even larger.

The aperture of the monolithic mirror telescope that has been built is 8 m, so we stipulate that the aperture of the relay mirror in four-mirror Nasmyth systems I and II is less than or equal to 8 m.

We stipulate that the major diameter of the 45° elliptical plane mirror in four-mirror Nasmyth systems I and II is ⩽4 m.

(2) The strip lensm in S-ADC is developed and has been initially successful. The experts responsible for the development think that there is no problem with manufacturing.

(3) The calculation in the example in this section shows that the rays go through a whole lensm strip and two lensm strips (i.e., for a point light source, light arriving at the surface of the S-ADC has or has no seam). In these two situations, their spot diagrams are similar, and the EE80D values are very close. It dispels the dilemma about the image quality of the rays going through two lensm strips. Moreover, the calculation of all subsequent spot diagrams only needs to be performed according to the rays going through a whole lensm strip.

(4) In sect. 4, the example is a 12-m aperture telescope with an FOV of 1.5°. However, for spectroscopic surveys, the etendue is directly proportional to the square of the FOV, and an FOV as small as 1.5° is not competitive. This example is only used to illustrate the design method, and we do not actually adopt it. The above spots are generated by Zemax and manually integrated into one diagram.

### 4.2 An example design for a 12-m aperture pure reflecting optical system telescope with an FOV of 1.5°

The four-mirror Nasmyth system I is adopted for the spectroscopic survey telescope as an example, as described in sect. 2. This is a pure reflecting system. Multiobject optical fiber spectroscopic observation can be performed on one side of the Nasmyth platform. The SYZ relay mirror (M3) and plane mirror (M4) are used to image the internal Cassegrain focus and direct the light to the Nasmyth focus. Figure 8 illustrates the optical layout of the 12-m spectroscopic survey telescope, with the exit pupil located very close to M4. The main optical parameters of this telescope are as follows:

(1) telescope diameter: 12 m (segmented);
(2) focal ratio: 4.0;
(3) FOV: Φ1.5°;
(4) effective focal length: 48 m;
(5) wavelength range: 0.36-1.8 μm;
(6) because the requirement of the whole FOV is the same for the multiobject spectroscopic survey, the four-mirror reflecting system is optimized under such a requirement.

Table 1 summarizes the optical parameters for the pure reflecting optical system.

The image quality of the pure reflecting system is excellent, and the EE80D is encircled in 0.1 arcsec. Figure 9 illustrates the spot diagrams. The vignetting value is 19%, which comes from light blocking, and the linear obstruction ratio is 42%.

### 4.3 ADC design for the spectroscopic survey focus

Suppose the altitude of the observation site is 4200 m, and the atmospheric pressure is 620 mbar. The ADC design is based on the S-ADC (Figure 10). A single strip has a length of 1344 mm, width of 192 mm, central thickness of 9.6 mm + 9.6 mm, and inner plane inclination angle of 3.9°. The edge thickness of the strip is 16.1 mm + 3.1 mm. The glass materials are LLF1 and N-BAK2 (the refractive indices and internal transmittance are shown in Tables 2 and 3[1)]), which have similar refractive indices at 1.06 μm wavelength. There

**Table 1** Main optical parameters of the 12-m four-mirror Nasmyth system I with an FOV of 1.5°[a)]

| Mirror | *D* (mm) | CR (mm) | *d* (mm) | Conic | 4th order term | 6th order term | 8th order term |
|---|---|---|---|---|---|---|---|
| Primary | Φ12000 | $-2.4\times10^{4}$ | 9430.815 | −0.87215 | – | – | – |
| Secondary | Φ2896 | $-1.112\times10^{4}$ | 11036.653 | – | $6.237\times10^{-13}$ | $-1.341\times10^{-20}$ | $3.553\times10^{-28}$ |
| Tertiary | Φ3982 | −8548 | 6260.301 | −0.41694 | – | $-5.666\times10^{-24}$ | – |
| Fold | 2852×1877 (ellipse) | flat | 7209.500 | – | – | – | – |
| Focal surface | Φ1262 | 7408 | – | – | – | – | – |

a) *D* is the clear aperture, CR is the curvature radius, and *d* is the distance to the next surface along the optical axis.

1) Schott Optical Glass Datasheets Website: https://www.schott.com/.

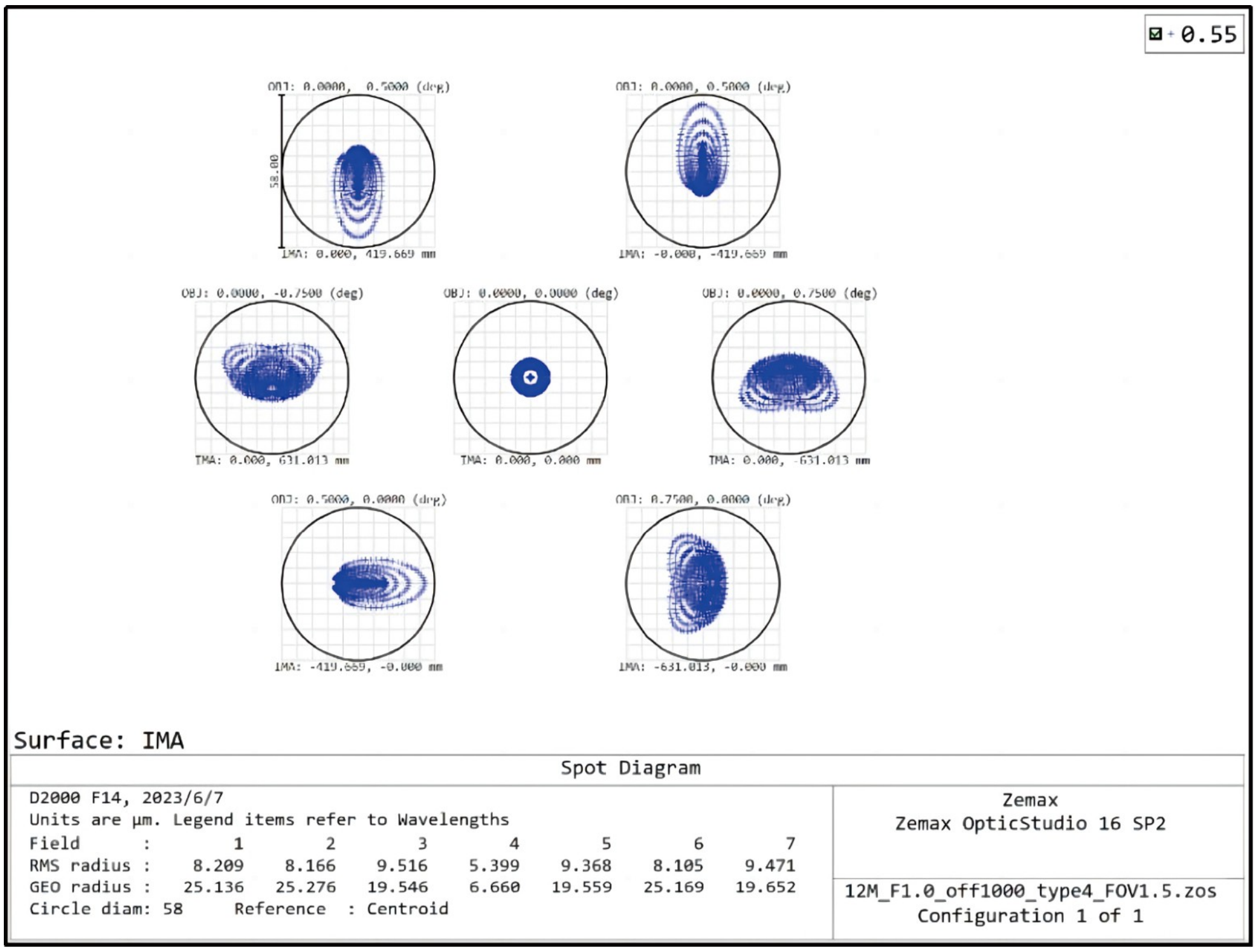

**Figure 9** (Color online) Spot diagrams of the typical reflecting system (without atmospheric dispersion). The circle diameters correspond to 58 μm (0.25 arcsec). The spot diagrams are not strictly symmetrical because M4 is set to decenter to reduce light blocking.

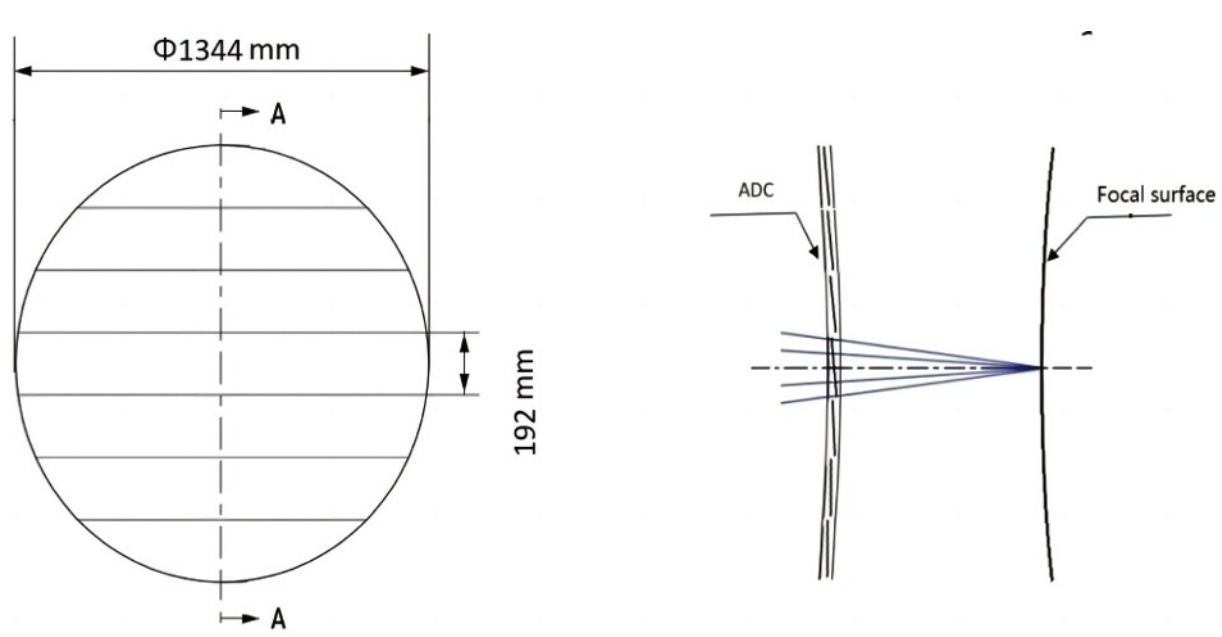

**Figure 10** Layout of S-ADC. Left: front view of the S-ADC. Right: the AA profile map of the S-ADC, with the thickness being doubled for clearer display.

**Table 2** Refractive indices and internal transmittances of N-BAK2

| Wavelength (nm) | Refractive indices | Wavelength (nm) | Internal transmittance (10-mm thickness) |
|---|---|---|---|
| 1970.1 | 1.51871 | 1970 | 0.940 |
| 1060.0 | 1.52919 | 1060 | 0.999 |
| 852.1 | 1.53234 | 700 | 0.998 |
| 656.3 | 1.53721 | 660 | 0.998 |
| 632.8 | 1.53806 | 620 | 0.998 |
| 587.6 | 1.53996 | 580 | 0.998 |
| 486.1 | 1.54625 | 460 | 0.997 |
| 435.8 | 1.55117 | 436 | 0.997 |
| 365.0 | 1.56221 | 365 | 0.994 |
| 334.1 | 1.56971 | 334 | 0.963 |

is a very small difference in expansion coefficient between these two types of glass, which is beneficial for cementing. The curvature center of the S-ADC surface is located at the same position as the center of the exit pupil; thus, the aberration, which is introduced by the S-ADC, is minimal. Tables 4 and 5 list the parameters of the S-ADC.

The following design results are obtained: The EE80D values in different FOVs at zenith distances of 60° are shown in Table 6. Table 7 summarizes the image quality. It can be observed in Figures 11, 12 and Table 6 that the shapes with the seam and seamless image spots are similar, and EE80D is very similar. The image spots in the sky area with a zenith distance smaller than 60° are smaller, so for 45°, 30°, 15°, and 0° zenith distances, we only show the results where the rays go through a whole lensm strip.

For the star at 60° zenith distance, the diameter of the star-illuminated area of the central FOV on the S-ADC is 165 mm, which is 184 mm for the star at the edge FOV. At 0° zenith distance, the distance between the edge of the S-ADC and the edge of the focal surface is 75 mm in the chief ray direction.

Let us follow up with some discussion of the system.

(1) For this 12-m, FOV 1.5° telescope, the maximum angle between the chief ray and the normal of the focal surface is

**Table 3** Refractive indices and internal transmittances of LLF1

| Wavelength (nm) | Refractive indices | Wavelength (nm) | Internal transmittance (10-mm thickness) |
|---|---|---|---|
| 1970.1 | 1.52354 | 1970 | 0.930 |
| 1060.0 | 1.53470 | 1060 | 0.998 |
| 852.1 | 1.53845 | 700 | 0.999 |
| 656.3 | 1.54457 | 660 | 0.998 |
| 632.8 | 1.54566 | 620 | 0.998 |
| 587.6 | 1.54814 | 580 | 0.999 |
| 486.1 | 1.55655 | 460 | 0.998 |
| 435.8 | 1.56333 | 436 | 0.998 |
| 365.0 | 1.57932 | 365 | 0.992 |
| 334.1 | 1.59092 | 334 | 0.920 |

**Table 4** Parameters of S-ADC

| Surface | Curvature radius (mm) | Thickness (mm) | Glass | Tilt angle | Width of one lensm strip (mm) |
|---|---|---|---|---|---|
| 1 | −6146.4 | 9.6 | LLF1 | −3.90° | 192 |
| 2 | −6156.0 | 9.6 | N-BAK2 | | |
| 3 | −6165.6 | variable | – | | |
| Focal surface | 7407.6 | – | – | | |

**Table 5** Positions of S-ADC and focal surface

| Zenith distance (°) | Distance between surface 3 of the ADC and the focal surface (mm) | Distance between M4 and the focal surface (mm) |
|---|---|---|
| 0 | 0.66 | 7216.26 |
| 15 | 69.62 | 7216.42 |
| 30 | 200.45 | 7216.75 |
| 45 | 369.91 | 7217.11 |
| 60 | 661.01 | 7217.81 |

9.8°, and for a 12-m, FOV 2.5° telescope, such an angle will be 14.7°. We think that this problem can be resolved by modified the present optical fiber positioning system. For example, method one: each fiber positioning system is inserted into the based body on the focal surface, at an angle to make the optical fiber aims the chief ray, and adding a focusing mechanism (for different observation objects, the fiber need to be moved a little for focusing). Method two: each fiber positioning system is perpendicularly inserted into the based body, which lets the fiber end always on the focal surface, and a small prism is put on it to deflect light into the direction of the focal surface radius, making the fiber aims the chief ray.

(2) However, it has been calculated that if the entire thickness were to be increased by 5 mm and the thinnest point was to become 5.6 mm, the image quality EE80D would be increased by 0.05 arcsec, which would still be a quite feasible and acceptable result.

**Table 6** EE80D in different FOVs at 60° zenith distance

| Field of view (°) | EE80D (arcsec) (in Figure 12 top) | EE80D (arcsec) (in Figure 12 bottom) |
|---|---|---|
| (0, 0) | 0.36 | 0.36 |
| (0, 0.5) | 0.32 | 0.33 |
| (0, 0.75) | 0.31 | 0.31 |
| (0, −0.5) | 0.35 | 0.35 |
| (0, −0.75) | 0.35 | 0.36 |
| (0.5, 0) | 0.31 | 0.31 |
| (0.75, 0) | 0.31 | 0.33 |

**Table 7** Summary of maximum EE80D at 0°-60° zenith distances

| Zenith distance (°) | EE80D (arcsec) |
|---|---|
| 0 (without atmospheric dispersion) | 0.10 |
| 0 (with ADC) | 0.22 |
| 15 | 0.19 |
| 30 | 0.22 |
| 45 | 0.26 |
| 60 | 0.36 |

(3) If the length of a lensm strip is >1.8 m (the current limit optical glass size), two half-length strips can be used. Even if all strips consist of half-strips, i.e., with the addition of a perpendicular seam in the S-ADC, the negative effect is insignificant. According to Figure 11 and Table 6, the image quality is almost the same whether the star-illuminated area has a seam or not.

(4) Thus, the aberration, which is introduced by the S-ADC, will be minimal.

### 4.4 Coudé system of the 12-m spectroscopic survey telescope with an FOV of 1.5°

The second SYZ relay mirror was adopted to image the Nasmyth focus to the coudé focus of the 12-m spectroscopic survey telescope with an FOV of 1.5°, while M2 was moved by −7.36 mm to achieve good performance. Without considering atmospheric dispersion, the image quality EE80D is less than the diameter of 0.1 arcsec over a full FOV of Φ8 arcmin, and the diffraction is limited at 1 μm wavelength within Φ3 arcmin. The optical layout of the coudé system is shown in Figure 13. The image spot diagram is shown in Figure 14, in which the outer circle represents 1 arcsec. In this design, a deformable mirror (DM) is conjugated with 139 m above the primary mirror, and the tip-tilt mirror is included for ground-layer adaptive optics correction. The inclination angle of the DM is 18° with an aperture of 1230 mm × 1300 mm. The coudé system's focal ratio is F/28.9.

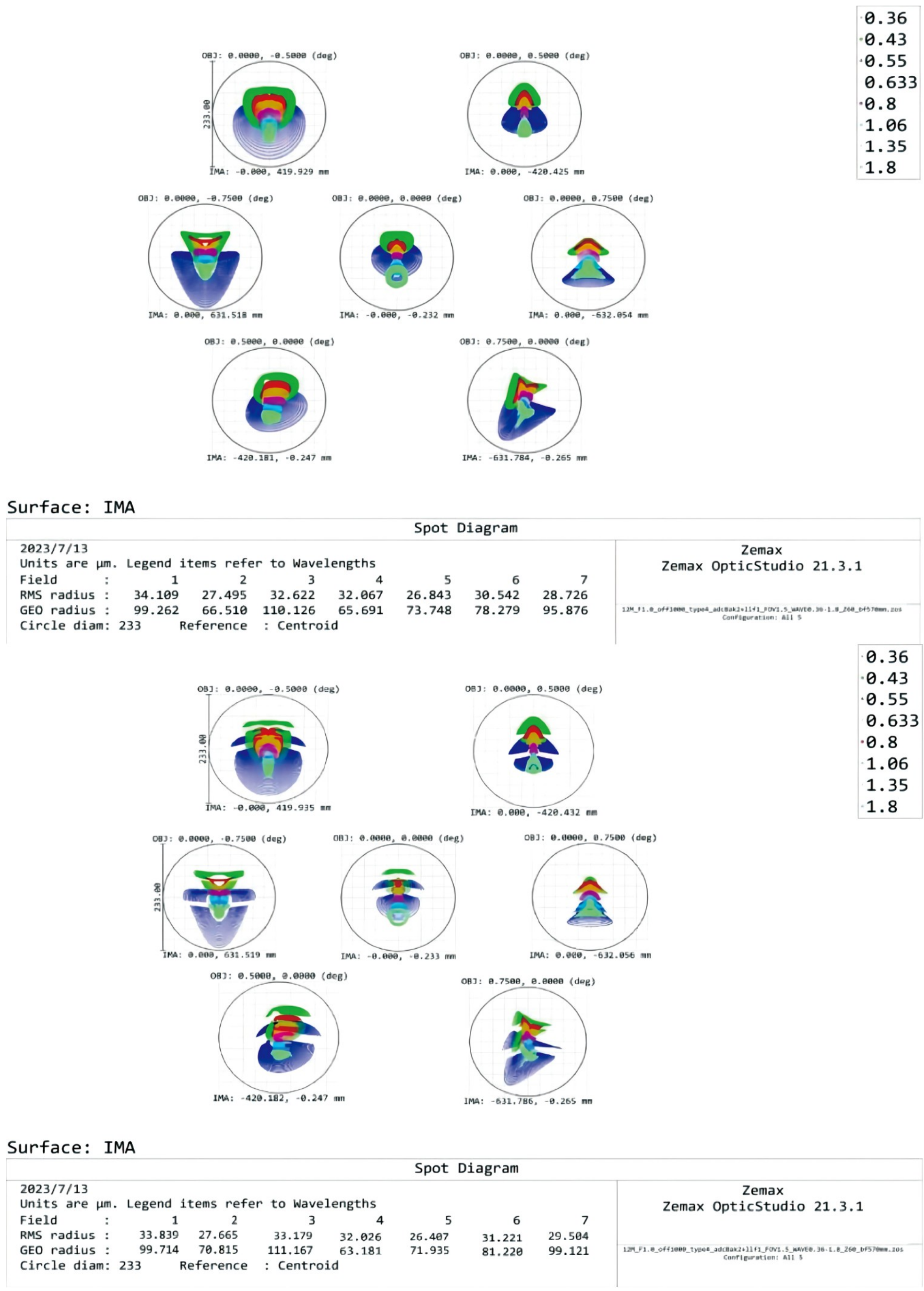

**Figure 11** (Color online) Top: spot diagram with S-ADC at 60° zenith distance. The rays go through a whole lensm strip. Bottom: spot diagram with S-ADC at 60° zenith distance where the rays go through two lens-prism strips, and the seam equally divides the illuminated area. The two diagrams are from the same focal surface and the positions of the lensm strips. The circle diameters correspond to 233 µm (1 arcsec). In this paper, Zemax is used for all optical system optimizations.

When the telescope is operating over a broad band from visible to near-infrared light, atmospheric dispersion should be corrected. In this design, a pair of counter-rotation lensms can be inserted at ~3.2 m before the focal surface. The glass materials for the lensms are LLF1 and N-BAK2. The lensm diameter is 863 mm, and the in-between surface tilt angle is 2.74°/3.24°. The image quality EE80D values are less than diameters of 0.105, 0.113, 0.14, 0.19, and 0.28 arcsec under different zenith distances of 0°, 15°, 30°, 45°, and 60°, respectively, over a full FOV of Φ8 arcmin. The image spots are shown in Figure 15.

Each of the 12-m telescopes with FOVs of 2° and 2.5° could also have a coudé system like in Figure 13.

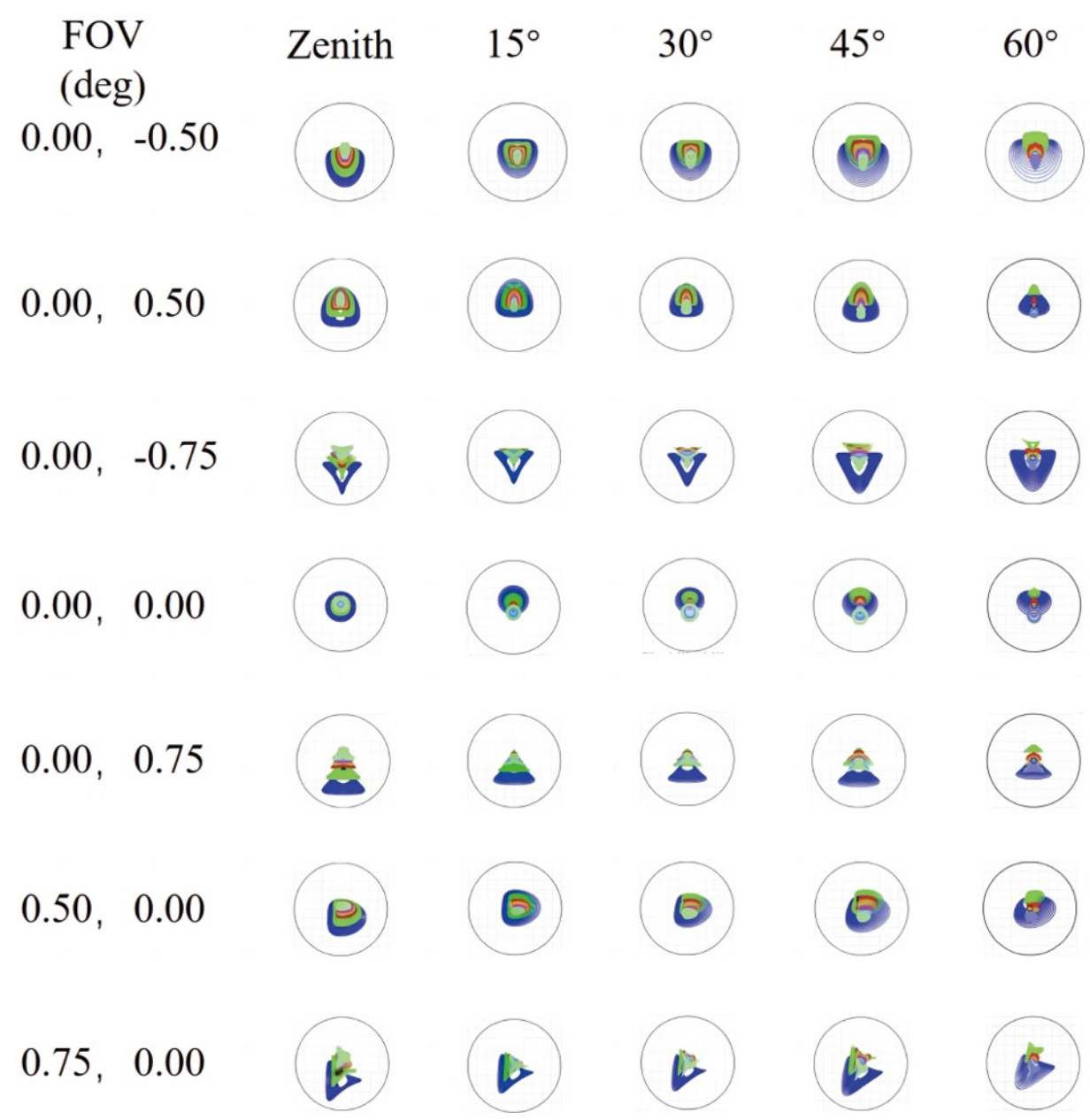

**Figure 12** (Color online) Spot diagram of the 12-m spectroscopic survey telescope with an FOV of 1.5° and ADC. The wavelength range is 0.36-1.8 μm, the site altitude is 4200 m, and the zenith distance is 0°-60°. The circle diameters correspond to 1 arcsec (233 μm). The maximum EE80D is 0.36 arcsec. The above spots are generated by Zemax and manually integrated into one diagram.

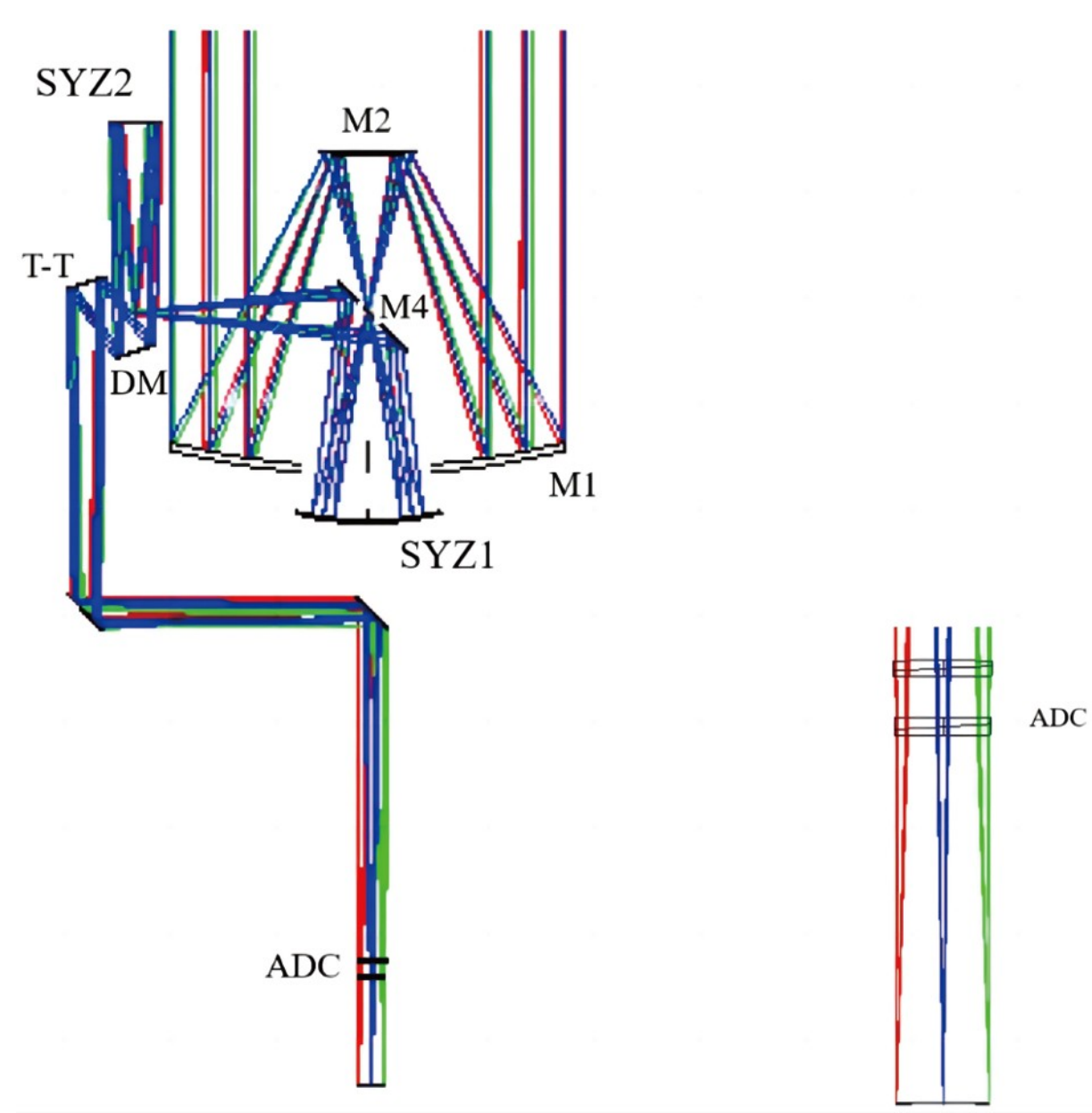

**Figure 13** (Color online) Coudé system optical layout of the 12-m spectroscopic survey telescope.

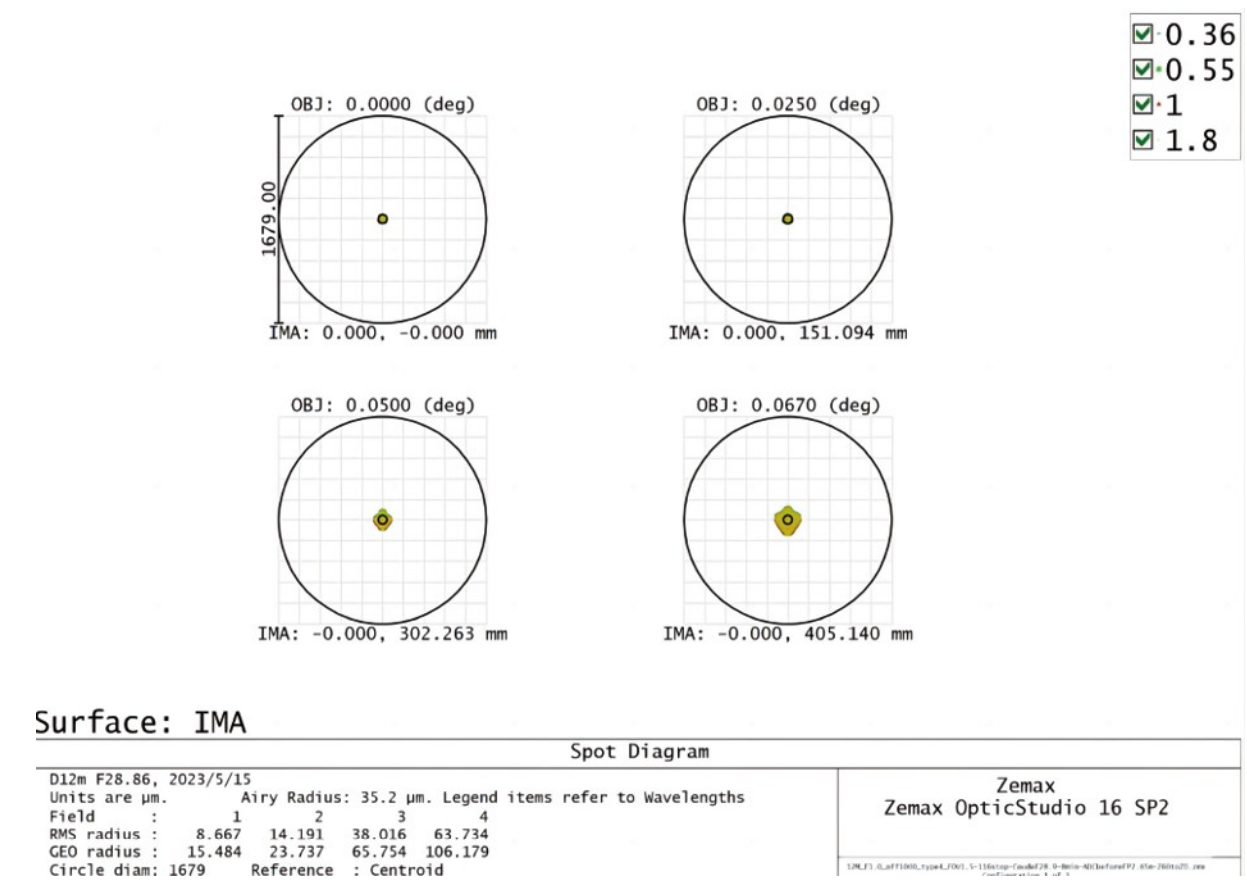

**Figure 14** (Color online) Image spot of the coudé system without considering the atmospheric dispersion effect.

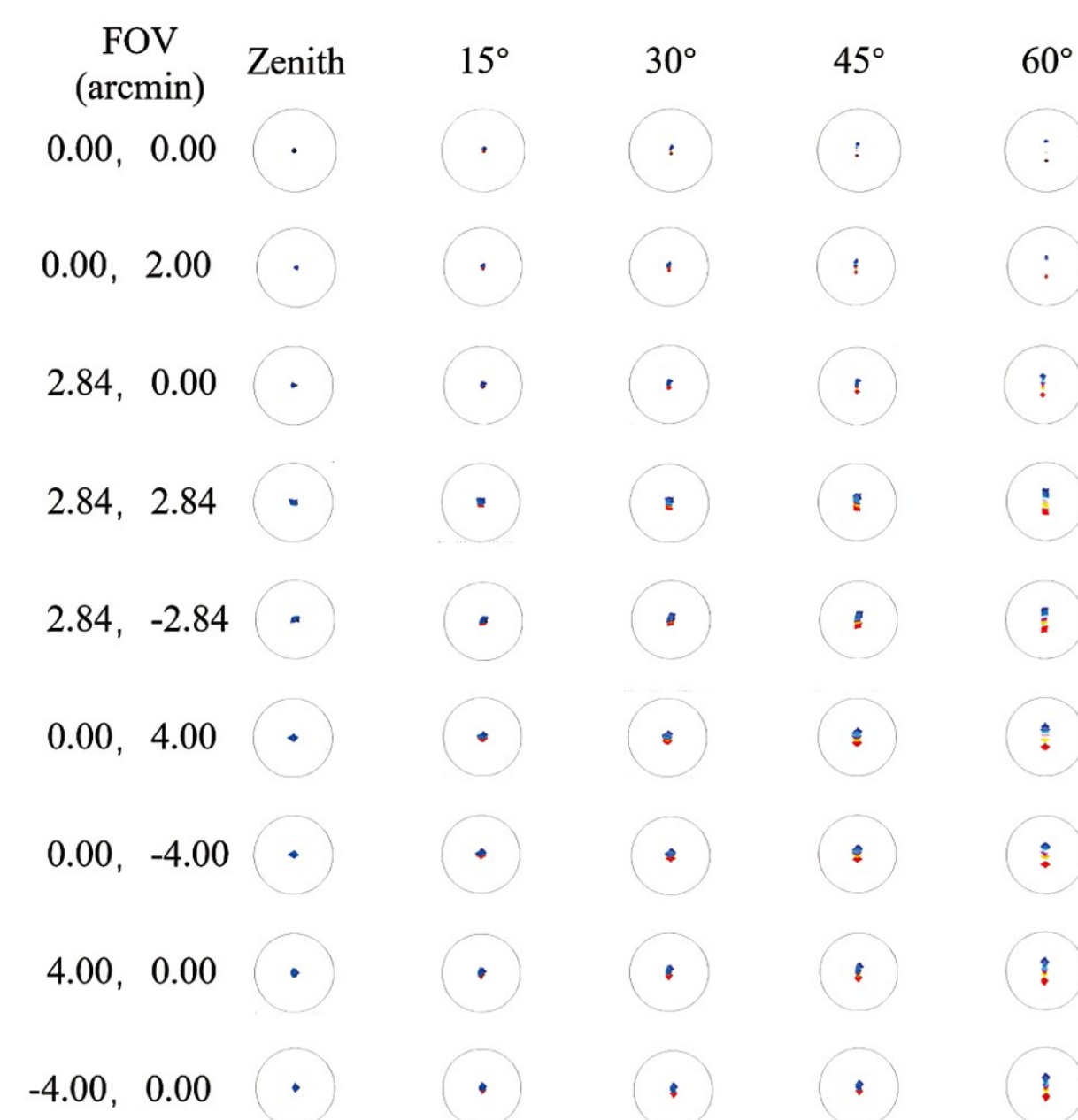

**Figure 15** (Color online) Spot diagram of the 12-m spectroscopic survey telescope coudé system with an FOV of 8 arcmin, wavelength range of 0.36-1.8 μm, site altitude of 4200 m, and zenith distance of 0°-60°. The circle diameters correspond to 1 arcsec (233 μm). The maximum EE80D is 0.28.

## 5 12-m four-mirror Nasmyth system I (or II) with an FOV of 2°

Figure 16 illustrates the optical layout of the system. The main optical parameters of this system are as follows:

(1) telescope diameter: 12 m (segmented);

(2) focal ratio: 4.0;

(3) FOV: Φ2.0°;

(4) effective focal length: 48 m;

(5) wavelength range: 0.36-1.8 μm.

Tables 8 and 9 show the optical parameters for the pure reflecting optical system and ADC. The image quality is excellent and presented in Table 10. The spot diagrams are shown in Figures 17 and 18. The maximum angle at the focal surface edge between the chief ray and the normal of the focal surface is 11.7°. Referring to sect. 2.3, which uses the four-mirror Nasmyth system I, a corresponding four-mirror

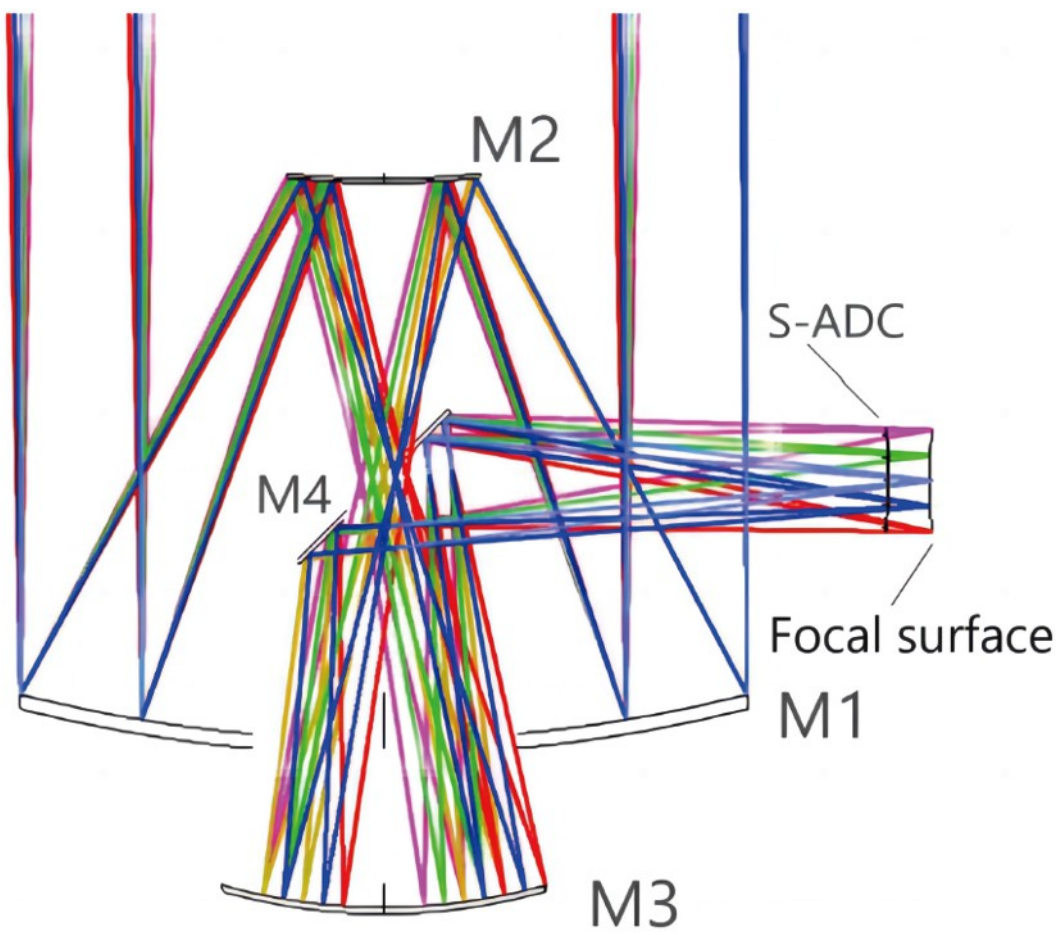

**Figure 16** (Color online) 12-m four-mirror Nasmyth system I spectroscopic survey telescope with an FOV of 2°.

Nasmyth system II can be obtained.

## 6 12-m four-mirror Nasmyth system II with an FOV of 2.5°

For the 12-m four-mirror Nasmyth I system with an FOV of 2.5°, the fold plane mirror M4 will be more than 4 m, so only the four-mirror Nasmyth system II shown in Figure 19 can be adopted. The main optical parameters of this telescope are as follows:

(1) telescope diameter: 12 m (segmented);
(2) focal ratio: 4.0;
(3) FOV: Φ2.5°;
(4) effective focal length: 48 m;
(5) wavelength range: 0.36-1.3 μm.

Tables 11 and 12 show the optical parameters for the pure reflecting optical system and ADC. The image quality is presented in Table 13. The spot diagrams are shown in Figures 20 and 21. From Table 11, for the 12-m FOV of the 2.5° telescope, the focal surface diameter is 2112 mm, and about 54000 optical fibers can be accommodated (assuming the diameter of the zone for one optical fiber is 8 mm).

Some discussions are presented as follows.

(1) The seeing of a world-class observatory site is about 0.70 arcsec, and the corresponding EE80 is about 1.05 arcsec. For a 12-m telescope with FOV 2.5°, its maximum EE80D is 0.65 arcsec. The square root of the sum of the squares of 1.05 and 0.65 is equal to 1.23 arcsec. Dividing 1.23 by 1.05 equals 1.17, the image spot increases by about 17%, but increasing the FOV from 1.5° and 2° to 2.5° gives a much greater benefit, so we finally choose the system with FOV 2.5°.

In the ESO white paper "SpecTel: A 10-12 meter class spectroscopic survey telescope", an 11.4-m spectroscopic survey telescope with FOV 2.5°, the RMS diameters of image spots are from 0.3 to 0.7 arcsec for zenith distance 0°-55° and wavelength 0.36-1.20 μm. For our 12-m telescope with FOV 2.5°, the RMS diameters of image spots are also from 0.3 to 0.7 arcsec, but most of them are less than 0.5 arcsec for zenith distance 0°-60° and wavelength 0.36-1.30 μm.

(2) We set up a two-layer large-area Nasmyth platform [17,18,31] and place a large number of spectrometers and other instruments at the platform, with the advantages of short-length optical fiber and less energy loss. It is necessary and normal to adopt a distance of 5.2-meter from the edge of the telescope tube to the Nasmyth focus.

(3) If one thinks that the Nasmyth platform is too large, and so the rotation radius of the telescope is too large. The observation room can be built inside the telescope's foundation pier. Many spectrometers and other instruments are installed in the foundation pier. we propose two alternated specific schemes: 1) Do not change the optical system, but a fold plane mirror could be installed near the mirror tube, to turn the focal surface to the lower part of the telescope fork-arm, then to let the optical fiber bundle into the observation rooms. 2) Change the power (curvature radius) of the relay mirror to reduce the distance from the edge of the telescope tube to the Nasmyth focus from 5.2 to 3 m, and then the focal ratio will be changed from 4 to 3.5, and the light loss will increase by about 4%.

(4) From Table 11, the diameter of the relay mirror (tertiary) is 5.954 m. According to the issue (1) in sect. 4.1, this

**Table 8** Main optical parameters of the 12-m four-mirror Nasmyth I (or II) system with an FOV of 2°[a)]

| Mirror | $D$ (mm) | CR (mm) | $d$ (mm) | Conic | 4th order term | 6th order term | 8th order term |
|---|---|---|---|---|---|---|---|
| Primary | Φ12000 | $-2.4\times10^{4}$ | 9200.000 | −0.85347 | – | – | – |
| Secondary | Φ3218 | $-1.3\times10^{4}$ | 11966.800 | – | $-4.5270\times10^{-14}$ | $-8.7599\times10^{-20}$ | $1.9343\times10^{-28}$ |
| Tertiary | Φ5343 | −9794 | 7047.881 | −0.41085 | – | $-3.5880\times10^{-24}$ | – |
| Fold | 3560 × 2400[b)] (ellipse) | flat | 9000.000 | – | – | – | – |
| Focal surface | Φ1686 | 7640 | – | – | – | – | – |

a) $D$ is the clear aperture, CR is the curvature radius, and $d$ is the distance to the next surface along the optical axis; b) the inner bore is 1620 mm × 920 mm (ellipse). In the corresponding four-mirror Nasmyth system II, the fold plane mirror only equals M4 inner bore 1620 × 920.

**Table 9**  Parameters of S-ADC of the 12-m four-mirror Nasmyth I (or II) system with an FOV of 2°

| Surface | Curvature radius (mm) | Thickness (mm) | Glass | Tilt angle | Width of one lensm strip (mm) |
|---|---|---|---|---|---|
| 1 | −8320 | 10 | LLF1 | −3.90° | 200 |
| 2 | −8330 | 10 | N-BAK2 | | |
| 3 | −8340 | variable | – | | |
| Focal surface | 7640 | – | – | | |

**Table 10**  Summary of EE80D of the 12-m four-mirror Nasmyth I (or II) system with an FOV of 2°, wavelength range of 0.36-1.8 μm, site altitude of 4200 m, and zenith distance of 0°-60°

| Zenith distance (°) | EE80D (arcsec) |
|---|---|
| 0 (without atmospheric dispersion) | 0.18 |
| 0 (with ADC) | 0.33 |
| 15 | 0.26 |
| 30 | 0.27 |
| 45 | 0.32 |
| 60 | 0.39 |

12-m FOV 2.5° telescope can be scaled up into a 16-m FOV 2.5° telescope. There are eight telescopes built around the world in the early 2000s, all of which have 8-m monolithic primary mirrors. For this case, we believe it is feasible to use an 8-m monolithic mirror for the relay mirror. In addition, the relay mirror can be mounted on the Nasmyth platform rather than on the telescope tube there is no rotation and no gravity change.

(5) In the four-mirror Nasmyth system II, one Nasmyth focus is blocked by the large relay mirror. Here is a brief introduction of a solution we came up with: this focus can be mainly used for fine observation and infrared observation, generally with 20 arcmin of FOV. Considering Figure 19 with the 12-m optical system and FOV 2.5°, if we open a hole in the center of the relay mirror with a diameter of about 1.3 m, replace the 2000 × 1140 ellipse plane mirror with a small ellipse plane mirror, and on its right side about 0.6 m away place a round plane mirror of about 2.7 m in diameter, which will reflect the light coming from the left-side relay mirror back through the hole of this relay mirror to the Nasmyth focus. Moreover, the plane mirror of approximately 2.7 m is roughly conjugated with the primary mirror, which is suitable as a GLAO DM.

From Table 11, the diameter of the focal surface is 2112 mm, and about 50000 optical fibers can be accommodated (assuming the diameter of the zone for one optical fiber is 8 mm).

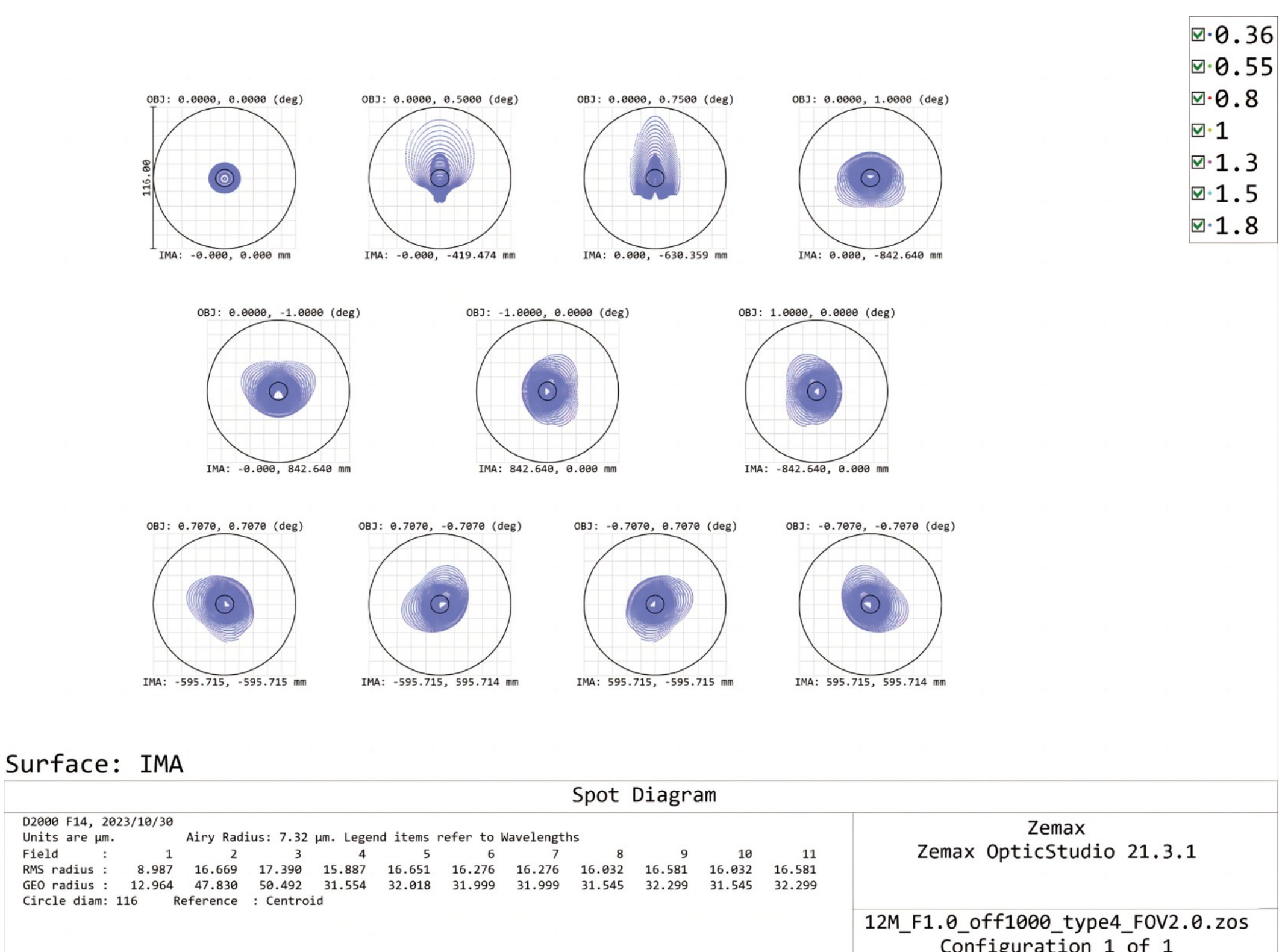

**Figure 17**  (Color online) Spot diagrams of the 12-m four-mirror system Nasmyth I with an FOV of 2° (without atmospheric dispersion). The circle diameters correspond to 0.50 arcsec (116 μm). The spot diagrams are not strictly symmetrical because M4 is set to decenter to reduce light blocking. The maximum EE80D is 0.18 arcsec in an FOV of 2°. In the central FOV of 8 arcmin, the maximum EE80D is 0.1 arcsec.

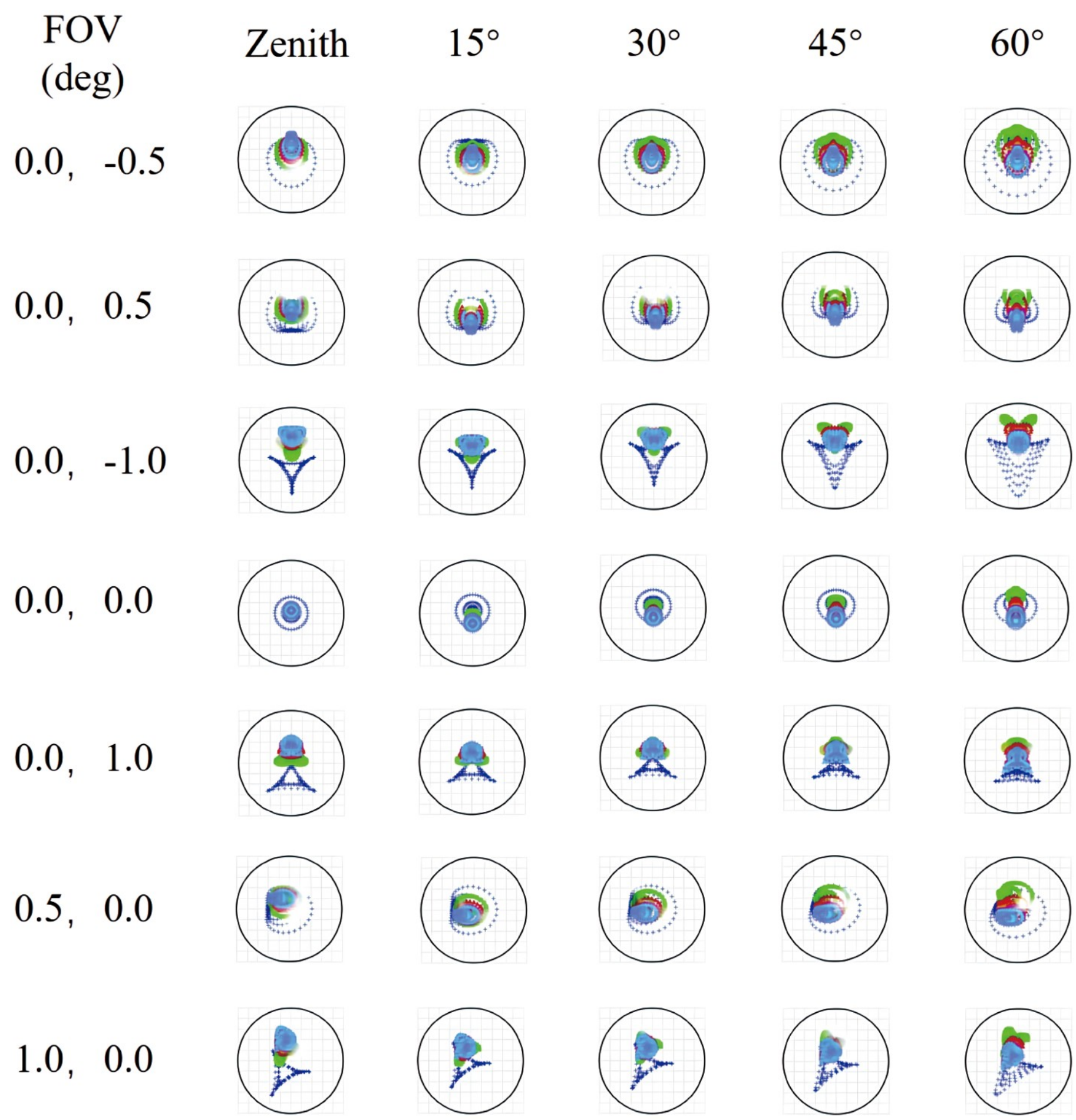

**Figure 18** (Color online) Spot diagram of the 12-m four-mirror system Nasmyth I with an FOV of 2° with ADC. The wavelength range is 0.36-1.8 μm, the site altitude is 4200 m, and the zenith distance is 0°-60°. The circle diameters correspond to 1.2 arcsec (279 μm). The maximum EE80D is 0.39 arcsec.

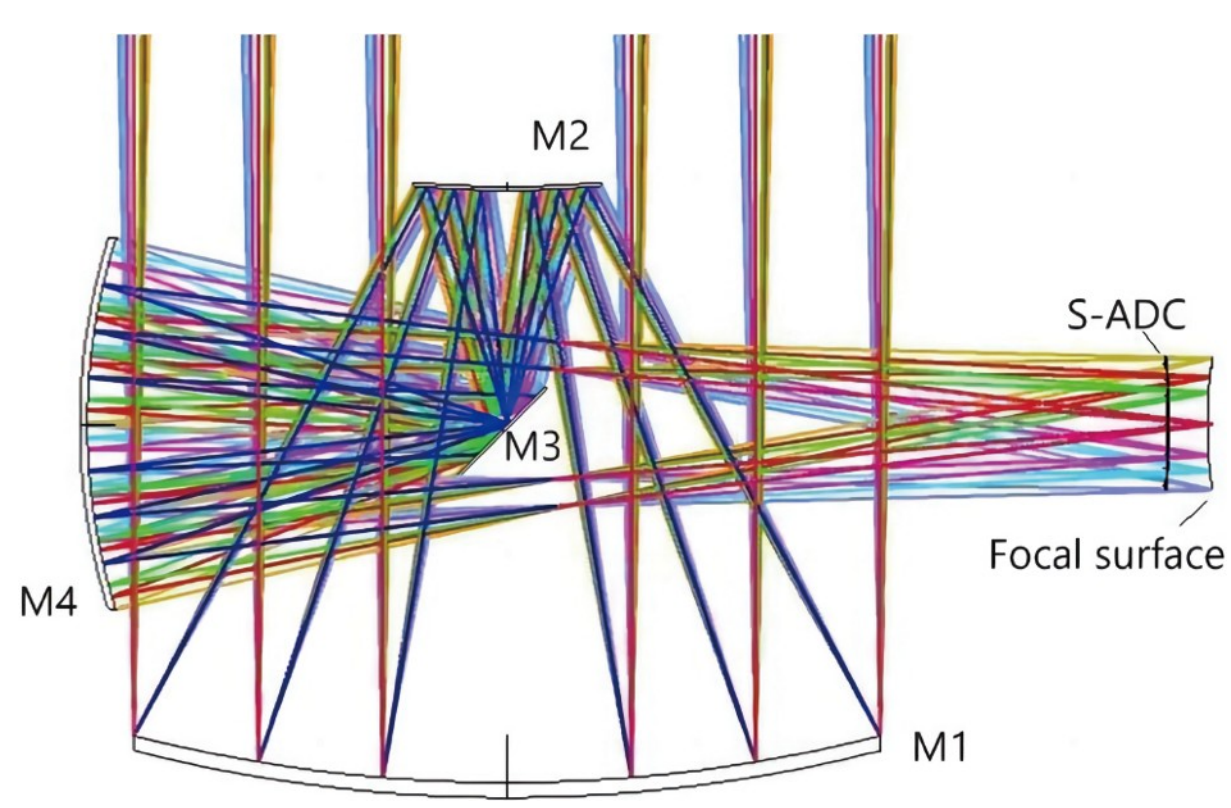

**Figure 19** (Color online) 12-m four-mirror Nasmyth II system with an FOV of 2.5°.

## 7 Discussion and conclusions

### 7.1 Comparison of the optical system capabilities for the four telescope configurations

In Table 14, we listed the main performance and etendue values for the telescope configurations, including in our paper with a diameter of 12 m and an FOV of 2.5°, and scaled it up into a telescope configuration with a diameter of 16 m and an FOV of 2.5°; the ESO 11.4-m concept [27]; and the 14-m telescope designed by Barden et al. [32]. where etendue 1 is equal to the telescope's light-gathering area (in $m^2$) times the FOV (in $deg^2$), and etendue 2 is equal to the telescope's light-gathering area (in $m^2$) times the focal surface area (in $m^2$).

Regarding the configuration in our paper, with a diameter of 12 m and an FOV of 2.5°, etendue 1 is similar to that of the ESO concept but has an exceeded result of etendue 2. In our configuration with a diameter of 12 or 16 m and an FOV of 2.5°, both etendues 1 and 2 obviously exceed Barden's design.

### 7.2 Taking the same aperture of 14 m and comparison of optical system capabilities with Barden et al. [32]

When the telescope aperture is 14 m, comparing our optical system capabilities with those of Barden et al.:

(1) For our FOV 2.5° configuration, etendues 1 and 2 are

**Table 11** Main optical parameters of the 12-m four-mirror Nasmyth system II with an FOV of 2.5°[a)]

| Mirror | *D* (mm) | CR (mm) | *d* (mm) | Conic | 4th order term | 6th order term | 8th order term | 10th order term |
|---|---|---|---|---|---|---|---|---|
| Primary | Φ12000 | $-2.4\times10^{4}$ | 9500.000 | −0.85619 | – | – | – | – |
| Secondary | Φ3218 | $-1.5\times10^{4}$ | 3750.000 | – | $5.2625\times10^{-13}$ | $-1.8631\times10^{-20}$ | $1.9590\times10^{-27}$ | $-2.6412\times10^{-34}$ |
| Fold | 2000 × 1140 (ellipse) | flat | 6749.716 | – | – | – | – | – |
| Tertiary | Φ5954 | 9818 | 17998.997 | – | $-5.4752\times10^{-14}$ | $-4.6121\times10^{-22}$ | $2.3019\times10^{-30}$ | $-7.9767\times10^{-38}$ |
| Focal surface | Φ2112 | 6681 | – | – | – | – | – | – |

a) *D* is the clear aperture, CR is the curvature radius, and *d* is the distance to the next surface along the optical axis.

**Table 12** Parameters of S-ADC of the 12-m four-mirror Nasmyth system II with an FOV of 2.5°

| Surface | Curvature radius (mm) | Thickness (mm) | Glass | Tilt angle | Width of one lensm strip (mm) |
|---|---|---|---|---|---|
| 1 | −10600 | 10 | LLF1 | −3.90° | 200 |
| 2 | −10610 | 10 | N-BAK2 | | |
| 3 | −10620 | variable | – | | |
| Focal surface | 6681 | – | – | | |

**Table 13** The 12-m four-mirror Nasmyth system II[a)]

| Zenith distance (°) | EE80D (arcsec) |
|---|---|
| 0 (without atmospheric dispersion) | 0.43 |
| 0 (with ADC) | 0.65 |
| 15 | 0.48 |
| 30 | 0.50 |
| 45 | 0.52 |
| 60 | 0.62 |

a) Summary of maximum EE80D. The wavelength range is 0.36-1.3 μm with ADC, the site altitude is 4200 m, and the zenith distance is 0°-60°.

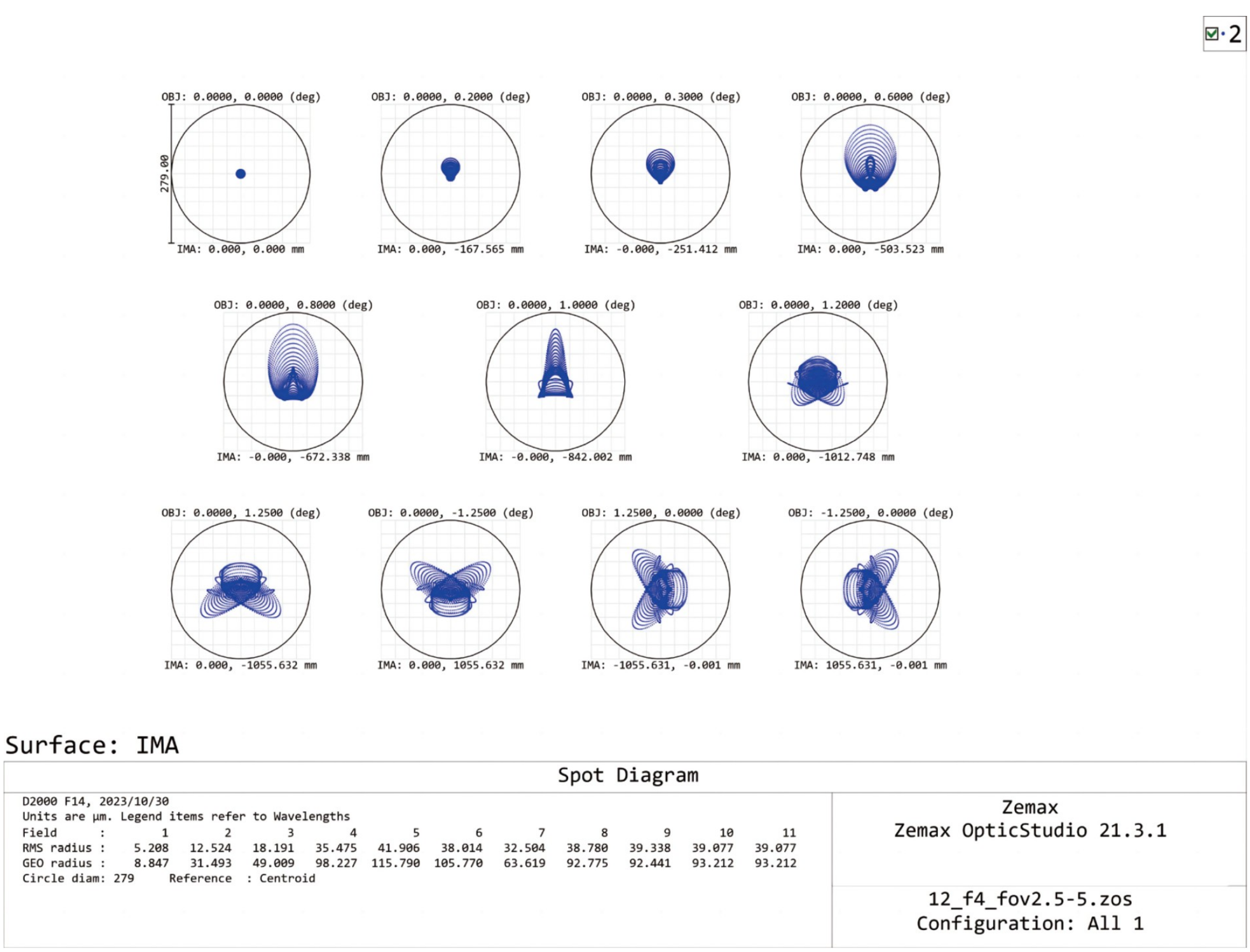

**Figure 20** (Color online) Spot diagrams of the 12-m four-mirror Nasmyth II system with an FOV of 2.5° (without atmospheric dispersion). The circle diameters correspond to 1.2 arcsec (279 μm). The spot diagrams are not strictly symmetrical because M4 is set to decenter to reduce light blocking. The maximum EE80D is 0.43 arcsec in an FOV of 2.5°. In the central FOV of 8 arcmin, the maximum EE80D is 0.08 arcsec.

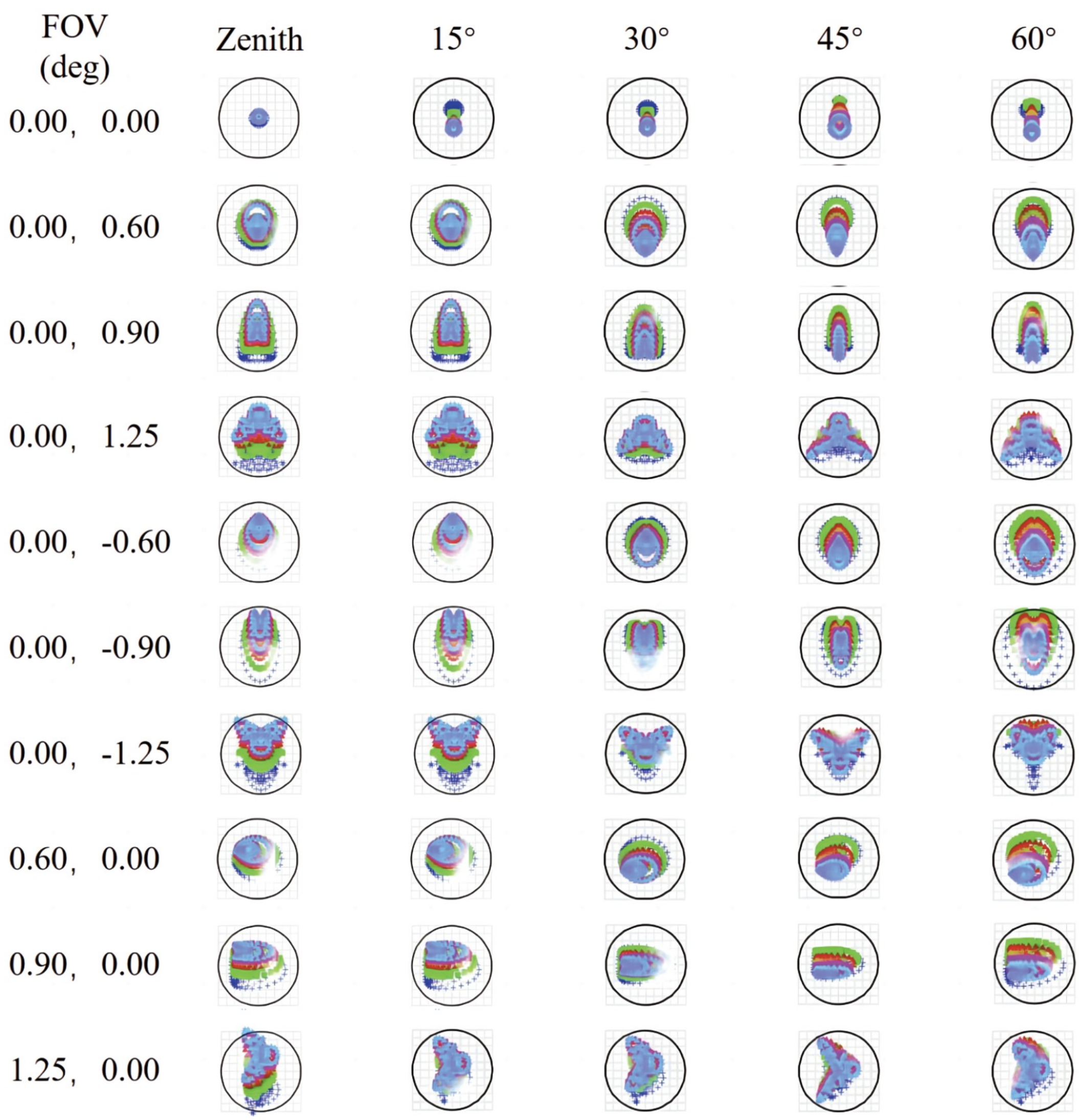

**Figure 21** (Color online) Spot diagram of the 12-m four-mirror Nasmyth system II with an FOV of 2.5° with ADC. The wavelength range is 0.36-1.3 μm, the site altitude is 4200 m, and the zenith distance is 0°-60°. The circle diameters correspond to 1.2 arcsec (279 μm). The maximum EE80D is 0.65 arcsec.

605 and 587, respectively.

(2) For the FOV 1.5° configuration of Barden et al. [32], etendues 1 and 2 are 223 and 214, respectively.

Obviously, for the same telescope aperture of 14 m, our configuration with an FOV of 2.5° is significantly better than that of Barden et al. [32].

### 7.3 Conclusions

The optical systems of the spectroscopy survey telescopes are generally composed of two parts: the reflective mirrors and the refractive lenses, which combine to eliminate aberration and atmospheric dispersion. The telescope aperture is limited by the size of the lens material. Currently the lens material is 1.8 m maximum. The idea of this paper is: a pure reflecting optical system is designed, which should have an aperture and a field of view (FOV) both as large as possible and with excellent image quality, then a S-ADC is added only to correct the atmospheric dispersion for the spectroscopy survey.

Because the S-ADC relaxes the optical glass size restriction, a 16-m aperture with 2.5° FOV spectroscopic survey telescope can be obtained. By the other means, it is difficult to realize such a telescope optical system. From paragraphs 7.1 and 7.2, one can see that the etendues 1 and 2 of our configuration with 16-m aperture are the greatest in the world now. There are two foci in our configurations, one of which is a pure reflecting optical system used for the fine and infrared observations while the atmospheric dispersion is negligible. A subsequent coudé system has been designed with excellent image quality for high resolution spectroscopy and Integral Field Units (IFU).

Except the four-mirror Nasmyth system I, II used in this paper, and LAMOST, we also researched another pure re-

**Table 14** Comparison of etendue for the four spectroscopic survey telescope configurations[a)]

| Telescope configs | DT (m) | Ratio | SL ($m^2$) | FOV (°) | Ω ($deg^2$) | DF (m) | SF ($m^2$) | Etendue 1 | Etendue 2 |
|---|---|---|---|---|---|---|---|---|---|
| Our Nasmyth I | 12 | 0.80 | 90.8 | 2.5 | 4.91 | 2.112 | 3.50 | 446 | 318 |
| Our Nasmyth II Max. Dia. | 16 | 0.80 | 161 | 2.5 | 4.91 | 2.816 | 6.23 | 793 | 1005 |
| ESO concept [27] | 11.4 | 0.86 | 87.89 | 2.5 | 4.91 | 1.43 | 1.61 | 431 | 141 |
| Barden et al. [32] | 14 | 0.82 | 126 | 1.5 | 1.77 | 1.47 | 1.70 | 223 | 214 |

a) DT is the aperture diameter of the telescope. Ratio is the observable effective area of the telescope and the area of the primary mirror of the telescope. SL is the area of light collection. DF is the diameter of the focal surface. SF is the area of the focal surface. Our Nasmyth I (or II) means four-mirror Nasmyth I (or II) in this article.

flecting systems, such as the Korsch [20] and Meinel et al. [24], Rumsey-Lemaitre [3]. These systems are three non-plane mirror only, and the final focus is outside the secondary mirror.

*This work was supported by the National Key R&D Program of China (Grant No. 2022YFA1603001). We thank Professor Gerard Lemaitre, the famous astronomer and optical scientist and one of the two pioneers of monolithic mirror active optics, for his valuable discussions, suggestions, and encouragement on this paper. We also thank Professor Ming Liang, the famous astronomical optical expert who designed the optical systems of DESI, DKIST and LSST (one of the two designers) for the USA and the work mentioned above for China, for his valuable discussions on this paper. We thank Professor Gongbo Zhao, a distinguished astrophysicist and the deputy director of NAOC, for his encouragement and support. We thank Ms. Xiaojie Jiao and Ms. Tingting Liu very much for helping us process some text.*

**Conflict of interest** The authors declare that they have no conflict of interest.

1 S. Wang, D. Su, and Q. Hu, Proc. SPIE **2199**, 341 (1994).
2 S. Wang, D. Su, Y. Chu, X. Cui, and Y. Wang, Appl. Opt. **35**, 5155 (1996).
3 G. R. Lemaitre, *Astronomical Optics and Elasticity Theory: Active Optics Methods* (Springer-Verlag, Heidelberg, 2009).
4 X. Q. Cui, Y. H. Zhao, Y. Q. Chu, G. P. Li, Q. Li, L. P. Zhang, H. J. Su, Z. Q. Yao, Y. N. Wang, X. Z. Xing, X. N. Li, Y. T. Zhu, G. Wang, B. Z. Gu, A. L. Luo, X. Q. Xu, Z. C. Zhang, G. R. Liu, H. T. Zhang, D. H. Yang, S. Y. Cao, H. Y. Chen, J. J. Chen, K. X. Chen, Y. Chen, J. R. Chu, L. Feng, X. F. Gong, Y. H. Hou, H. Z. Hu, N. S. Hu, Z. W. Hu, L. Jia, F. H. Jiang, X. Jiang, Z. B. Jiang, G. Jin, A. H. Li, Y. Li, Y. P. Li, G. Q. Liu, Z. G. Liu, W. Z. Lu, Y. D. Mao, L. Men, Y. J. Qi, Z. X. Qi, H. M. Shi, Z. H. Tang, Q. S. Tao, D. Q. Wang, D. Wang, G. M. Wang, H. Wang, J. N. Wang, J. Wang, J. L. Wang, J. P. Wang, L. Wang, S. Q. Wang, Y. Wang, Y. F. Wang, L. Z. Xu, Y. Xu, S. H. Yang, Y. Yu, H. Yuan, X. Y. Yuan, C. Zhai, J. Zhang, Y. X. Zhang, Y. Zhang, M. Zhao, F. Zhou, G. H. Zhou, J. Zhu, and S. C. Zou, Res. Astron. Astrophys. **12**, 1197 (2012).
5 X. Xing, C. Zhai, H. Du, W. Li, H. Hu, R. Wang, and D. Shi, Proc. SPIE **3352**, 839 (1998).
6 D. Su, P. Jia, and G. Liu, Mon. Not. R. Astron. Soc. **419**, 3406 (2012).
7 H. Bai, D. Q. Su, M. Liang, S. A. Shectman, X. Y. Yuan, and X. Q. Cui, Res. Astron. Astrophys. **21**, 132 (2021).
8 D. Su, Astron. Astrophys. **156**, 381 (1986).
9 D. Su, and M. Liang, Proc. SPIE **628**, 479 (1986).
10 M. Liang, and D. Su, in Very large telescopes and their instrumentation: Proceedings of a ESO Conference on Very Large Telescopes and their Instrumentation, European Southern Observatory (ESO), Garching, 1988, p. 237.
11 Y. Wang, and D. Su, Astron. Astrophys. **232**, 589 (1990).
12 R. N. Wilson, *Reflecting Telescope Optics I* (Springer-Verlag, Berlin, 1996).
13 T. Agócs, M. Balcells, C. R. Benn, D. C. Abrams, and D. C. Infantes, Proc. SPIE **7735**, 773563 (2010).
14 G. Liu, and X. Yuan, Acta Astron. Sin. **50**, 224 (2005).
15 P. Doel, M. J. Sholl, M. Liang, D. Brooks, B. Flaugher, G. Gutierrez, S. Kent, M. Lampton, T. Miller, and D. Sprayberry, Proc. SPIE **9147**, 91476R (2014).
16 D. Su, M. Liang, X. Yuan, H. Bai, and X. Cui, Mon. Not. R. Astron. Soc. **460**, 2286 (2016).
17 D. Su, M. Liang, X. Yuan, H. Bai, and X. Cui, Mon. Not. R. Astron. Soc. **469**, 3792 (2017).
18 X. Cui, Y. Zhu, M. Liang, D. Su, X. Yuan, Z. Hu, H. Bai, and B. Gu, Proc. SPIE **10700**, 107001 (2018).
19 D. Su, B. Zhou, and X. Yu, Sci. China A **33**, 454 (1990).
20 D. Korsch, Appl. Opt. **11**, 2986 (1972).
21 L. Goldberg, Sky and Telescope **56**, 383 (1978).
22 A. B. Meinel, M. P. Meinel, H. Ningshen, H. Qiqian, and P. Chunhua, Appl. Opt. **19**, 2670 (1980).
23 A. B. Meinel, and M. P. Meinel, Appl. Opt. **19**, 2683 (1980).
24 A. B. Meinel, and M. P. Meinel, *Current Trends in Optics* (Invited Papers from the ICO-12 Meeting, Graz, Austria, 1981), F. T. Arecchi, and F. R. Aussenegg, eds. (Taylor and Francis Ltd., London, 1981), p. 40.
25 D. Su, L. Shao, and M. Liang, Opt. Acta **29**, 1237 (1982).
26 L. Pasquini, B. Delabre, R. S. Ellis, and T. de Zeeuw, Proc. SPIE **9906**, 99063C (2016).
27 L. Pasquini, B. Delabre, R. S. Ellis, J. Marrero, L. Cavaller, and T. de Zeeuw, Proc. IAU Sym. **334**, 242 (2017).
28 W. Saunders, P. R. Gillingham, G. Smith, S. Kent, and P. Doel, Proc. SPIE **9151**, 91511M (2014).
29 P. R. Gillingham, and W. Saunders, Proc. SPIE **9151**, 61 (2014).
30 W. Saunders, and P. R. Gillingham, Proc. SPIE **9906**, 38 (2016).
31 D. Su, Y. Wang, and X. Cui, Proc. SPIE **5489**, 429 (2004).
32 S. Barden, M. Baril, and D. Jones, Proc. SPIE **12182**, 121822I (2022).